\documentclass[twocolumn,pra,amsmath,amssymb,floatfix,superscriptaddress,showpacs]{revtex4-1}

\usepackage{amsmath,amssymb,graphicx,mathtools,bbold}
\usepackage{color}
\usepackage{bbm}
\usepackage[caption=false]{subfig}

\usepackage[mathscr]{euscript}

\DeclarePairedDelimiter\bra{\langle}{\rvert} 
\DeclarePairedDelimiter\ket{\lvert}{\rangle} 
\DeclarePairedDelimiterX\braket[2]{\langle}{\rangle}{#1 \delimsize\vert #2}

\newcommand{\dd}{\mathrm{d}}

\begin{document}
\title{Non-Equilibrium Quantum Spin Dynamics from Classical Stochastic Processes}
 \author{S. De Nicola}
\affiliation{IST Austria, Am Campus 1, 3400 Klosterneuburg, Austria}
\author{B. Doyon}
\affiliation{Department of Mathematics, King's College London, Strand,
  London WC2R 2LS, United Kingdom} \author{M. J. Bhaseen}
\affiliation{Department of Physics, King's College London, Strand,
  London WC2R 2LS, United Kingdom}

\begin{abstract}
Following on from our recent work, we investigate a stochastic approach to non-equilibrium quantum spin systems. We show how the method can be applied to a variety of physical observables and for
different initial conditions. We provide exact formulae of broad applicability for the time-dependence of expectation values and
correlation functions following a quantum quench in terms of averages over classical stochastic processes. We further explore the behavior of the classical stochastic variables in the presence of dynamical quantum phase transitions, including results for their distributions and correlation functions. We provide details on the numerical solution of the associated stochastic differential equations, and examine the growth of fluctuations in the classical description. 
We discuss the strengths and limitations of the current implementation of the stochastic approach and the potential for further development.\end{abstract}

\maketitle
\section{Introduction}

The experimental realization of isolated quantum many-body systems
\cite{newtonsCradle,singleSpin1,blattRoos2012,langen2015,reviewColdAtoms2015,Gross2017}
has led to intense theoretical interest in their unitary
time-evolution \cite{polkovnikov2011,eisert2015}. The study of quantum
quenches \cite{calabreseCardy2005,calabreseCardy2006} has provided
fundamental insights into their non-equilibrium behavior, including
the absence of thermalization in low-dimensional integrable systems
\cite{newtonsCradle,intro2016} and the role of the Generalized Gibbs
Ensemble (GGE) \cite{rigol2006,rigol2007,rigol2008}. This has
stimulated the development of new theoretical tools and methodologies,
ranging from the quench action approach \cite{caux2013,caux2016} to
recent applications of hydrodynamics
\cite{BhaseenDoyonLucasSchalm2015,
  DoyonLucasSchalmBhaseen2015,bernard2016,doyon2016,bertini2016}. This
has been complemented by significant advances in numerical simulation
techniques \cite{Cazalilla2002,Vidal2003,Vidal2004,White2004,Garcia-Ripoll2006,Schollwock2011}. The theoretical prediction of dynamical quantum phase transitions (DQPTs) \cite{heyl2013,heyl2018}, which occur as a function of time, has been recently confirmed using trapped ions \cite{jurcevic2017}. These experiments provide a new set of tools for exploring the time-resolved dynamics of quantum many-body systems using paradigmatic spin Hamiltonians.

Recently, a theoretical approach to quantum spin systems has
emerged, based on a mapping to classical stochastic processes
\cite{hoganChalker,galitski,ringelGritsev,stochasticApproach}. The
procedure begins by decoupling the exchange interactions between spins
using Hubbard--Stratonovich transformations. This yields an exact
description in terms of independent quantum spins, where the effect of
interactions is represented by Gaussian distributed stochastic
fields. Quantum expectation values are then expressed as classical
averages over these stochastic fields.  In recent work
\cite{stochasticApproach}, we showed that this approach could be used
to calculate the expectation values of time-dependent quantum
observables, including the experimentally measurable Loschmidt rate
function and the magnetization. We also verified that this approach
could handle both integrable and non-integrable models, including
those in higher dimensions. Here, we extend our previous work in a
number of directions, providing results for a broader range of
observables under different initial conditions.  We also present more
information on the stochastic approach itself and its numerical
implementation.
We also present new results on the dynamics of the classical
stochastic variables, including stochastic bounds on the
Loschmidt rate function.
For other recent work
exploring the connections between quantum and classical dynamics see
Refs~\cite{Polkovnikov2010,Ng2011,Wurtz2018}.

The layout of this paper is as follows. In Section
\ref{sec:disentanglement} we recall the principal steps involved in
the stochastic approach to quantum spin systems, adopting the
notations of Refs~\cite{ringelGritsev,stochasticApproach}. In Section
\ref{sec:stochastic} we show how quantum observables can be computed
in the stochastic formalism providing results of general applicability
for spin-$1/2$ systems. In Section \ref{sec:ising} we illustrate the
method by considering quenches in the quantum Ising model, in one and
two spatial dimensions. In Section \ref{sec:classical} we investigate
the relationship between DQPTs and the classical stochastic
variables. In Sections \ref{sec:fluctuations} and \ref{sec:cost}
  we discuss the strengths and limitations of the stochastic approach,
  exploring the growth of fluctuations in the classical variables and
  the computational cost of numerical simulations. We conclude in
  Section \ref{sec:conclusion}, summarizing our findings and
  indicating directions for future research. We also provide
appendices on the technical details of the stochastic approach and its
numerical implementation.

\section{Stochastic Formalism}\label{sec:disentanglement}
In this section we recall the principal steps involved in the
stochastic approach to quantum spin systems
\cite{hoganChalker,galitski,ringelGritsev,stochasticApproach}. Following
Refs~\cite{ringelGritsev,stochasticApproach}, we begin our discussion with a generic Heisenberg
Hamiltonian
\begin{equation}\label{eq:generalHamiltonian}
\hat{H} = - \sum_{ijab} \mathcal{J}_{ij}^{ab}\hat{S}_i^a \hat{S}_j^b -
\sum_{ja} h^a_j \hat{S}^a_j,
\end{equation}
where $i,j$ indicate lattice sites and $a,b$ label the spin
components. The spin operators satisfy the $su(2)$ commutation
relations $[\hat{S}^a_j,\hat{S}^b_k]=i \epsilon^{abc} \delta_{jk}
\hat{S}^c_k$, where $a,b,c \in \{x,y,z\}$, $\epsilon^{abc}$ is the
antisymmetric symbol and we set $\hbar=1$. The exchange interactions
$\mathcal{J}^{ab}_{ij}$ and the fields $h^a_j$ can, in general, be
time-dependent. Away from equilibrium, unitary dynamics under $\hat H$
is governed by the time-evolution operator
\begin{equation}\label{eq:timeEvolution}
\hat{U}(t_f,t_i) = \mathbb{T} \exp\left({ -i \int_{t_i}^{t_f} \mathrm{d} t\, \hat{H}(t)}\right), 
\end{equation}
where $t_i$ and $t_f$ denote the initial and final times, and
$\mathbb{T}$ denotes time-ordering. In general, the time-evolution
operator $\hat{U}(t_f,t_i)$ is non-trivial, due to the quadratic spin
interactions in $\hat H$, the non-commutativity of the spin operators,
and the time-ordering. However, some of
these difficulties can be circumvented in a two-step process. First,
the quadratic spin interactions in $\hat H$ can be decoupled exactly
using Hubbard--Stratonovich (HS) transformations. This leads to a
physically appealing description in terms of independent quantum spins
which are coupled via Gaussian distributed stochastic ``magnetic''
fields \cite{hoganChalker,ringelGritsev}. Second, the time-ordered
exponential in Eq.~(\ref{eq:timeEvolution}) can be recast as an
ordinary exponential; the HS decoupling renders the exponent linear in the
$su(2)$ generators, allowing a simpler parameterization via group theory
\cite{galitski,ringelGritsev}. This so-called {\em disentanglement
  transformation} \cite{ringelGritsev,stochasticApproach} can be
regarded as a judicious parameterization of the time-evolution
operator which takes advantage of the Lie algebraic structure
of the spin operators. In
Sections \ref{subsec:HST} and \ref{subsec:DT} we recall these two
steps in turn, before summarizing the resulting stochastic
differential equations (SDEs)
\cite{ringelGritsev,stochasticApproach}. In Section \ref{subsec:Ito}
we discuss the Ito form of these SDEs, which is useful for numerical
simulations.

\subsection{Hubbard--Stratonovich Transformation}
\label{subsec:HST}
As in Refs~\cite{hoganChalker,galitski,ringelGritsev,stochasticApproach}, the quadratic spin interactions can be
decoupled via a HS transformation \cite{stratonovich,hubbard} over
auxiliary variables $\varphi^a_j$. Trotter slicing \cite{trotter} the
exponential in Eq.~(\ref{eq:timeEvolution}) and applying the HS
transformation at each time slice (Appendix \ref{app:decoupling}) one obtains
\begin{equation}\label{eq:evOptr}
\hat{U}(t_f,t_i) = \mathbb{T} \int \mathcal{D} \varphi\,
\mathrm{e}^{-S[\varphi] + i \int_{t_i}^{t_f} \sum_{ja}
  \Phi^a_j(t^\prime) \hat{S}^a_j\,\mathrm{d}t^\prime},
\end{equation}
where we refer to
\begin{equation}\label{eq:noiseAction}
S[\varphi] = \sum_{ijab}\int_{t_i}^{t_f}
\frac{1}{4}(\mathcal{J}^{-1})_{ij}^{ab}\varphi^a_i(t^\prime)
\varphi^b_j(t^\prime)\,\mathrm{d}t^\prime,
\end{equation}
as the \textit{noise action}. Here, we define $\Phi_j^a \equiv
h^a_j+\varphi_j^a/\sqrt{i}$ and further denote
\begin{equation}
  {\mathcal D}\varphi \equiv \prod\limits_j {\mathcal D}
  \varphi_j^a,
  \end{equation}
  where ${\mathcal D}\varphi^a_j$ is the appropriately normalized
  integration measure for each HS variable $\varphi_j^a$. Introducing
  the change of variables $\varphi^a_i=\sum_{jb}O^{ab}_{ij}\phi^b_j$,
  where $O^{\rm T}\mathcal{J}^{-1} O/2 = \mathbb{1}$ and 
  $\mathbb{1}$ is the identity matrix, the noise action in
  Eq.~(\ref{eq:noiseAction}) can be recast in the diagonal form
\begin{equation}\label{eq:noiseActionDiagonal}
S[\phi] = \sum_{ia} \int_{t_i}^{t_f}  \frac{1}{2} \phi^a_i(t^\prime) \phi^a_i(t^\prime)\,\mathrm{d}t^\prime  ,
\end{equation}
where $\phi^a_i$ are real-valued Gaussian white noise variables
satisfying $\langle \phi^a_i(t) \rangle =0$, $\langle
\phi^a_i(t)\phi^b_j(t^\prime)\rangle
=\delta(t-t^\prime)\delta_{ij}\delta_{ab}$; see Appendix \ref{app:diag}. This yields a
probabilistic interpretation of Eq.~(\ref{eq:evOptr}) as an integral
over Gaussian weighted stochastic paths $\phi^a_i(t)$ \cite{hoganChalker,ringelGritsev}. The
time-evolution operator can thus be written in the form
\begin{equation}\label{eq:stochasticNotDisentangled}
\hat{U}(t_f,t_i)=\big\langle \mathbb{T} \mathrm{e}^{i \int_{t_i}^{t_f}  \sum_{ja} \Phi^a_j(t^\prime) \hat{S}^a_j\mathrm{d}t^\prime} \big\rangle_\phi  ,
\end{equation}
where $\Phi_j^a = h^a_j+\sum_{kb}O_{jk}^{ab}\phi_k^b/\sqrt{i}$ and $\langle
\dots \rangle_\phi$ denotes averaging with respect to the Gaussian
weight given by Eq.~(\ref{eq:noiseActionDiagonal}). Equivalently,
Eq.~(\ref{eq:stochasticNotDisentangled}) describes the time-evolution
of individual decoupled spins moving under the action of applied and
stochastic ``magnetic'' fields, $h_j^a(t)$ and ${\check
  h}_j^a(t)\equiv \varphi_j^a/\sqrt{i}=\sum_{kb}O_{jk}^{ab}\phi_k^b(t)/\sqrt{i}$, respectively. Although
the spins appear to be fully decoupled in the representation~(\ref{eq:stochasticNotDisentangled}), the effect of the interactions
is encoded in the fields $\check{ h}_j^a(t)$ via the matrix $O_{jk}^{ab}$. Each spin is governed by
an effective stochastic Hamiltonian
\begin{equation}\label{eq:effectiveH}
\hat{H}^s_j(t) = - \sum_a  \left( h_j^a(t)+ \check{h}_j^a(t)\right)\hat{S}^a_j. 
\end{equation}
In general, this is non-Hermitian, as the
stochastic fields $\check{h}^a_j$ may be complex valued.
Without loss of generality, in the remainder of this work we consider time-evolution over
the interval $[0,t]$ and 
set $\hat U(t)\equiv \hat U(t,0)$.
\subsection{Disentanglement Transformation}
\label{subsec:DT}
The time-evolution operator defined by
Eq.~(\ref{eq:stochasticNotDisentangled}) is still non-trivial due to
the time-ordering operation. However the decoupled exponential is now
linear in the spin operators, and can therefore be simplified using
group theory \cite{hoganChalker,galitski,ringelGritsev}. Specifically, one may rewrite the time-evolution
operator acting at a given site $j$ as
\begin{align}
  \label{eq:WNK}
\mathbb{T} \mathrm{e}^{i \sum_a \int_0^t \Phi^a_j(t^\prime)
  \hat{S}^a_j \mathrm{d}t^\prime} \equiv e^{\xi_j^+(t)
  \hat{S}_j^+} e^{ \xi_j^z(t) \hat{S}_j^z} e^{\xi_j^-(t) \hat{S}_j^-}
,
\end{align}
where the parameters $\xi_j^a(t)$ are referred to as \textit{disentangling
  variables} \cite{ringelGritsev}. This is also known as the
Wei--Norman--Kolokolov transformation \cite{weiNorman,kolokolov}. The
relationship between the disentangling variables $\xi_j^a(t)$ and the variables $\Phi_j^a(t)$ can be made more explicit by
differentiating Eq.~(\ref{eq:WNK}) with respect to time. This yields
\cite{ringelGritsev}
\begin{subequations}\label{eq:disentanglementGeneral}
\begin{align}
i \dot{\xi}^+_j &= \Phi^+_j + \Phi^z_j \xi^+_j - \Phi^-_j{\xi^+_j}^2 , \\
i \dot{\xi}^z_j &= \Phi^z_j - 2 \Phi^-_j \xi^+_j , \\
i \dot{\xi}^-_j &= \Phi^-_j \exp{\xi^z_j}, 
\end{align}
\end{subequations}
where the identity $\hat{U}(0)=1$ implies the initial conditions
$\xi^a_j(0)=0$ for all $j,a$. For completeness, we provide a
detailed derivation of these equations in
Appendix~\ref{app:disentanglement}. Alternative disentanglement
transformations, based on different group parameterizations, have also
been considered in the literature \cite{hoganChalker,ringelGritsev}.

Equations (\ref{eq:disentanglementGeneral}) may be regarded as
stochastic differential equations (SDEs) for the variables $\xi_j^a$,
due to the presence of the (additive and multiplicative) Gaussian
noise entering via $\Phi_j^a$ \cite{ringelGritsev}. Applying the disentanglement transformation (\ref{eq:disentanglementGeneral}) to the time-evolution operator
(\ref{eq:stochasticNotDisentangled}) one obtains
\cite{ringelGritsev,stochasticApproach}
\begin{equation}\label{eq:Udisent}
\hat{U}(t)=\langle \otimes_j \hat{U}^s_j(t) \rangle_\phi,
\end{equation} 
where we have defined on-site stochastic operators
\begin{equation}\label{eq:Usi}
\hat{U}^s_j(t) \equiv e^{\xi_j^+(t) \hat{S}_j^+}  e^{ \xi_j^z(t) \hat{S}_j^z}  e^{\xi_j^-(t) \hat{S}_j^-}.
\end{equation}
In general, this is a non-unitary operator, since the time-evolution
of each spin is governed by the non-Hermitian
Hamiltonian~(\ref{eq:effectiveH}).  Given a specific spin representation, $\hat{U}^s_j(t)$ can be written in matrix form. For
example, for spin-$\tfrac{1}{2}$, we may write $\hat{S}^a=\hat\sigma^a/2$ in terms of the Pauli matrices $\hat\sigma^a$, where $a\in\{x,y,z\}$. This yields
\begin{equation}\label{eq:onsiteU_expl}
\hat{U}^s_j(t)=   \begin{pmatrix}
   e^{\xi^z_j/2}  + e^{-\xi^z_j/2} \xi_j^+\xi^-_j &  e^{-\xi^z_j/2} \xi_j^+ \\ e^{-\xi^z_j/2} \xi_j^- & e^{-\xi^z_j/2} 
  \end{pmatrix}.
\end{equation}
The product form of the evolution operator~(\ref{eq:Udisent}) makes it
convenient for acting on spin states of interest; using this, the
quantum matrix elements of an operator $\hat{\mathcal{O}}(t)\equiv \hat U^\dagger(t) \hat{\mathcal O}\hat U(t)$ can be
expressed as the classical average of a function $f(\xi)$, over
realizations of the stochastic process:
\begin{align}\label{eq:generalObservable}
\langle \psi_{\rm F}|\hat{\mathcal{O}}(t)|\psi_{\rm I}\rangle = \langle f(\xi(t))\rangle_\phi.
\end{align}
Here, the function $f(\xi)$ depends on the disentangling variables
$\xi\equiv \{\xi^a_j\}$, and is determined by the observable
$\hat{\mathcal O}$, and the chosen initial and final states,
$|\psi_{\rm I}\rangle$ and $|\psi_{\rm F}\rangle$. In writing
(\ref{eq:generalObservable}), we consider operators $\hat{\mathcal O}$
without explicit time-dependence: in the Heisenberg picture their
time-evolution is determined solely by $\hat U(t)$. In the
Schr\"odinger picture, the matrix elements can be recast as
$\bra{\psi_{\rm F}(t)}\hat{\mathcal O}\ket{\psi_{\rm I}(t)}$, where
$\ket{\psi(t)}\equiv \hat U(t)\ket{\psi(0)}$. In Section
\ref{sec:stochastic} we will provide some explicit examples of the
quantum-classical correspondence (\ref{eq:generalObservable}), for
different observables and for different initial and final states.
\subsection{Ito Equations of Motion}
\label{subsec:Ito}
SDEs are defined by specifying a discretization scheme
\cite{kloeden}, with the most common choices being the Ito and
Stratonovich conventions. The SDEs (\ref{eq:disentanglementGeneral}) are initially in the
Stratonovich form. However, for numerical simulations, it is often
convenient to work with the Ito form of the SDEs, which are naturally
suited for discrete time-evolution. The Ito SDEs can be obtained by
including an extra drift term. However, in the case of an interaction
matrix $\mathcal{J}^{ab}_{ij}$ with vanishing diagonal elements, the
additional Ito drift term vanishes identically and the Ito and
Stratonovich SDEs coincide; see Appendix~\ref{app:itoStratonovich}.
The time-dependence of a function $f(\xi)$ corresponding to a physical
observable $\hat {\mathcal O}(t)$ can be found via Ito
calculus. For a generic Ito SDE written in the canonical form,
\begin{equation}
  \frac{d\xi_i^a}{dt}=A_i^a(\xi)+\sum_{jb}B_{ij}^{ab}(\xi)\phi_j^b ,
  \label{SDEs}
\end{equation}
one obtains
\begin{equation}\label{eq:EoMs}
 \dot{f} = \sum_{ia} \frac{\partial f}{\partial \xi^a_i}(A^a_i + \sum_{jb} B^{ab}_{ij} \phi^b_j) + \frac{1}{2} \sum_{ijab} \frac{\partial^2f}{\partial\xi_i^a \partial\xi_j^b} \sum_{ck} B^{ac}_{ik}  B^{bc}_{jk},
\end{equation}
as follows from Ito's lemma \cite{kloeden}. In principle, it is
possible to analytically average these SDEs with respect to the HS fields; in this approach, one obtains a system of ordinary differential equations (ODEs) \cite{ringelGritsev}.
However, as we discuss in Appendix~\ref{app:analyticalAveraging}, this
is formally equivalent to diagonalizing the Hamiltonian, whose matrix
dimension scales as ${\mathcal O}(2^N\times 2^N)$,
where $N$ is the total number of spins. Instead, it is more convenient
to numerically perform the average in (\ref{eq:generalObservable})
over independent realizations of the stochastic process. In this
approach, the number of stochastic variables $\xi_j^a$ that one needs
to simulate scales linearly with $N$. Moreover, the
independent runs can be readily parallelized. In Section
\ref{sec:stochastic}, we will provide exact stochastic formulae for a
variety of quantum observables that can be described in this way. We
will return to a more detailed discussion of the numerical aspects in
Sections \ref{sec:fluctuations} and \ref{sec:cost}.

\section{Quantum Observables}\label{sec:stochastic}
In order to illustrate the stochastic approach to non-equilibrium
quantum spin systems, we obtain below the classical formulae for a
range of quantum observables.

\subsection{Loschmidt Amplitude}
\label{subsec:loschmidt}
One of the simplest quantities to investigate in the stochastic
formalism is the Loschmidt amplitude $A(t)$. This is defined as the
amplitude for an initial state $\ket{\psi(0)}$ to return to itself
after unitary evolution \cite{heyl2013}:
\begin{equation}\label{eq:survivalAmplitude}
A(t)=\bra{\psi(0)}\hat{U}(t)\ket{\psi(0)}.
\end{equation}
In general,
$A(t)$ is expected to decay exponentially with the system size $N$. It
is therefore convenient to define the Loschmidt rate function
\begin{equation}
\lambda(t) \equiv -\frac{1}{N} \log |A(t)|^2.
\end{equation}
This plays the role of a dynamical free energy density, since $A(t)$
is analogous to a boundary partition function \cite{LeClair1995} that
is Wick-rotated to real time.
This connection led to the insightful prediction of dynamical quantum
phase transitions (DQPTs) occurring in $\lambda(t)$ as a function of
time \cite{heyl2013}. In the thermodynamic limit $N\rightarrow\infty$,
these transitions correspond to non-analyticities in $\lambda(t)$, and
often occur on quenching across a quantum critical point
\cite{heyl2013,heyl2018}. The existence of DQPTs has been recently
confirmed in experiment using trapped ions \cite{jurcevic2017}. This
experiment provides a realization of the quantum Ising model with $6$
to $10$ spins, interacting via tunable dipolar interactions. This
allows access to the time-resolved dynamics of an isolated quantum
spin system.

Here, we consider $A(t)$ for the generic Hamiltonian
(\ref{eq:generalHamiltonian}). In principle, this may contain long
range interactions as in the experiment \cite{jurcevic2017}, but this is not the
primary thrust of our investigation. For simplicity, we focus on
initial states of product form, $\ket{\psi(0)} = \otimes_i
\ket{\psi(0)}_i$. Parameterizing a generic superposition as
$|\psi(0)\rangle_i = a_i \ket{\uparrow}_i + b_i \ket{\downarrow}_i$,
where $\ket{\uparrow}$ and $\ket{\downarrow}$ refer to spin-up and
spin-down in the $\hat S_i^z$ basis, with $|a_i|^2 + |b_i|^2=1$, 
one obtains 
\begin{align}\label{eq:loschGen}
A(t) = \left\langle \prod_i  e^{-\frac{\xi^z_i}{2}} \left(|a_i|^2 e^{\xi^z_i}+(a_i  \xi^-_i +b_i ) (a^*_i  \xi^+_i +b_i^*)\right) \right\rangle_\phi.
\end{align}
In the special case of a fully-polarized initial state
$\ket{\Downarrow}$ with all spins down, corresponding to $a_i=0$ and
$b_i=1$, one obtains the result given in our previous work
\cite{stochasticApproach}. The result (\ref{eq:loschGen}) is more
general and allows consideration of spatially inhomogeneous initial
states. In Section \ref{sec:ising} we will discuss the numerical
evaluation of (\ref{eq:loschGen}) in the context of the quantum
Ising model, including domain wall initial conditions. For now, we
gather the stochastic formulae describing local observables.
\subsection{One-Point Functions}
The dynamics of a local observable $\hat{\mathcal O}$ is encoded in the time-dependent
expectation value
\begin{equation}\label{eq:onePoint}
\langle \hat{\mathcal{O}}(t) \rangle = \bra{\psi(0)} \hat{U}^\dag(t) \hat{\mathcal{O}} \hat{U}(t) \ket{\psi(0)}.
\end{equation}
In contrast to the Loschmidt amplitude (\ref{eq:survivalAmplitude}),
this involves two time-evolution operators. This can be addressed by 
two independent HS transformations over variables
$\phi\equiv\{ \phi_i^a \}$ and $\tilde{\phi}\equiv\{\tilde{\phi}_i^a\}$,
with associated disentanglement variables $\xi \equiv \{ \xi^a_i[
 \phi]\}$ and $\tilde{\xi} \equiv \{ \tilde{\xi}^a_i[
 \tilde{\phi}]\}$. 
For simplicity, we illustrate this in the
case where the observable $\hat{\mathcal{O}}$ of interest is a product
of $\hat{S}^z_i$ operators at different sites. This class of operators
includes the longitudinal magnetization as well as correlation
functions. For product initial states, the argument of the classical
average is factorized over the sites $i$. A given observable
$\hat{\mathcal{O}}$ can then be expressed in the stochastic language
by multiplying a set of on-site ``building blocks'', given in
Appendix~\ref{app:buildingBlocks}.
In this framework, local expectation values are expressed as averages
of functions of $\xi$ and $\tilde\xi$, describing the forwards and
backwards evolutions respectively.  For example, the dynamics of the
local magnetization for a system initialized in the state
$\ket{\psi(0)}=\ket{\Downarrow}$ is given by \cite{stochasticApproach}
\begin{equation}\label{eq:generalMagnetization}
\langle \hat{S}^z_i(t) \rangle = -\frac{1}{2}  \Big\langle e^{-\sum_j  \left( \frac{\xi^z_j+\tilde{\xi}^{z*}_j}{2} \right) } (1-\xi^+_i \tilde{\xi}^{+*}_i)  \prod_{j\neq i} (1+\xi^+_j \tilde{\xi}^{+*}_j)\Big\rangle_{\phi,\tilde{\phi}} .
\end{equation}
The structure of (\ref{eq:generalMagnetization}) is relatively straightforward. It consists of an exponential factor like
that in (\ref{eq:loschGen}), together with polynomial factors
$(1\pm\xi^+_j \tilde{\xi}^{+*}_j)$ for each site, where the minus sign is used for the chosen site $i$. As discussed in Section
\ref{sec:corr}, a similar structure also emerges in the evaluation of
correlation functions. Analogous results for $\langle S_i^\alpha(t)\rangle$ with $\alpha=x,y$ are given in Appendix~\ref{app:buildingBlocks}. 

\subsection{Equal-Time Correlation Functions}
\label{sec:corr}
Correlation functions of local operators can be computed in a similar
manner to that described above. For example,
the two-point function of the local magnetization $C_{ij}(t)\equiv \langle
\hat S^z_i(t) \hat{S}^z_j(t) \rangle$ is given by
\begin{widetext}
  \begin{equation}
    \label{eq:eqtimecorr}
\begin{split}
C_{ij}(t)= \frac{1}{4} \Big\langle & e^{-\sum_k \left( \frac{\xi^z_k+\tilde{\xi}^{z*}_k}{2} \right) } (1-\xi^+_i \tilde{\xi}^{+*}_i) (1-\xi^+_j \tilde{\xi}^{+*}_j)  \prod_{k\neq i, j} (1+\xi^+_k \tilde{\xi}^{+*}_k )\Big\rangle_{\phi,\tilde{\phi}},
\end{split}
\end{equation}
\end{widetext}
for quenches starting in $\ket{\Downarrow}$. The structure of
(\ref{eq:eqtimecorr}) mirrors that of (\ref{eq:generalMagnetization}),
where now there are two polynomial factors with minus signs, for
the chosen sites $i$ and $j$. This result is readily
generalized to arbitrary multi-point functions of the local
magnetization, starting in the state $\ket{\Downarrow}$; the sign of
the polynomial is negative for each factor of $\hat S_i^z$ in the
correlation function.
More generally, the expectation value of a product of local operators
starting from a product state can be decomposed into averages of
products of the elementary ``building blocks'' referred to above; see
Appendix \ref{app:buildingBlocks}. In the case of an initial
  state $\ket{\Downarrow}$, equations (\ref{eq:generalMagnetization})
  and (\ref{eq:eqtimecorr}) can be equivalently decomposed into a
  ``background'' factor $\prod_j e^{-\sum_j
    (\xi^z_j+\tilde{\xi}^{z*}_j)/2 }(1+\xi^+_j \tilde{\xi}^{+*}_j)$,
  for all the sites that are not involved in the observable, together
  with a multiplicative factor for each inserted local operator. This
  structure is reminiscent of the form of correlation functions
  obtained from the algebraic Bethe ansatz, see
  e.g. Ref.~\cite{Kitanine2005}, although the present results apply to
  both integrable and non-integrable problems.

\subsection{Dynamical Correlation Functions}
Dynamical correlation functions involving operators at different times can also be expressed in the stochastic formalism, by decoupling each of the time-evolution operators. For example, the two-time correlation function 
$C_{ij}(t,t^\prime) \equiv \langle\hat{S}^z_i(t)
\hat{S}^z_j(t^\prime)\rangle$ can be written as
\begin{align}
C_{ij}(t,t^\prime) = \bra{\psi(0)} \hat{U}^\dag(t) \hat{S}^z_i \hat{U}(t)  \hat{U}^\dag(t^\prime) \hat{S}^z_j \hat{U}(t^\prime) \ket{\psi(0)}.
\end{align}
Starting in the initial state $\ket{\psi(0)}=\ket{\Downarrow}$ and using $\hat{U}(t)  \hat{U}^\dag(t^\prime) = \hat{U}(t-t^\prime)$, one obtains
\begin{widetext}
  \begin{align}
    \label{eq:dynamcorr}
\begin{split}
C_{ij}(t,t^\prime) =\frac{1}{4} \Big\langle& e^{-\frac{1}{2} \sum_l \left({\xi^z_{1,l}(t)}^*+{\xi^z_{2,l}(t-t^\prime)}+{\xi^z_{3,l}(t^\prime)}\right)}\left[ ({\xi^-_{2,i}(t-t^\prime)} {\xi^+_{3,i}(t^\prime)}+1) \left({\xi^+_{2,i}(t-t^\prime)} {\xi^+_{1,i}(t)}^*-1\right)+{\xi^+_{3,i}(t^\prime)} e^{{\xi^z_{2,i}(t-t^\prime)}} {\xi^+_{1,i}(t)}^* \right] \\
 &  \hspace*{0.2cm} \left[ ({\xi^-_{2,j}(t-t^\prime)} {\xi^+_{3,j}(t^\prime)}-1) \left({\xi^+_{2,j}(t-t^\prime)} {\xi^+_{1,j}(t)}^*+1\right)+{\xi^+_{3,j}(t^\prime)} e^{{\xi^z_{2,j}(t-t^\prime)}} {\xi^+_{1,j}(t)}^* \right] \times
 \\
 & \hspace*{-1.4cm} \prod_{k \neq i,j}   \left({\xi^+_{1,k}(t)}^* \left({\xi^-_{2,k}(t-t^\prime)} {\xi^+_{2,k}(t-t^\prime)} {\xi^+_{3,k}(t^\prime)}+{\xi^+_{2,k}(t-t^\prime)}+{\xi^+_{3,k}(t^\prime)} e^{{\xi^z_{2,k}(t-t^\prime)}}\right)+{\xi^-_{2,k}(t-t^\prime)} {\xi^+_{3,k}(t^\prime)}+1\right) \Big\rangle_{\phi_1,\phi_2,\phi_3} .
\end{split}
\end{align}
\end{widetext}
Here we have introduced three sets of disentangling variables
$\xi_{1,2,3}$ which are functionals of three independent Gaussian
white noise fields $\phi_{1,2,3}$. Although the expression
(\ref{eq:dynamcorr}) is rather non-trivial, it is general to dynamical
correlations of arbitrary spin-$1/2$ Heisenberg models
(\ref{eq:generalHamiltonian}), without reference to integrability
  or dimensionality.
\subsection{Higher Dimensions}
A notable feature of the stochastic approach is that it applies to
systems in arbitrary dimensions. Due to the on-site character of the
stochastic time-evolution operators $\hat{U}^s_j(t)$, all of the formulae obtained above readily generalize to arbitrary
dimensions: the products simply extend over all the lattice sites. In
Section \ref{sec:ising} we will provide an example of this in the
context of the two-dimensional quantum Ising model.

\section{Quantum Ising Model}\label{sec:ising}
In order to illustrate how the stochastic method can be applied in
practice, we consider quantum quenches in the one-dimensional (1D)
quantum Ising model \cite{stochasticApproach}. The Hamiltonian is given by
\begin{align}\label{eq:isingH}
\hat{H}_I &= -J\sum\limits_{j=1}^N \hat{S}^z_j \hat{S}^z_{j+1} - \Gamma
\sum_{j=1}^N \hat{S}^x_j,
\end{align}
where $J>0$ is the ferromagnetic nearest neighbor exchange interaction
and $\Gamma$ is the transverse field. For simplicity, we consider
periodic boundary conditions with $\hat S^a_{N+1}=\hat S^a_1$. In
equilibrium, the model (\ref{eq:isingH}) exhibits a quantum phase
transition at $\Gamma=\Gamma_c\equiv J/2$ between a ferromagnetic (FM)
phase for $\Gamma< \Gamma_c$ and a paramagnetic (PM) phase for
$\Gamma>\Gamma_c$. Out of equilibrium, the dynamics of the Hamiltonian
(\ref{eq:isingH}) is encoded in the Ito SDEs
\begin{subequations}\label{eq:isingSDEs}
\begin{align}
-i \dot{\xi}^+_j &= \frac{\Gamma}{2} (1-{\xi^+_j}^2) + \xi^+_j \sum_k
O_{jk} \phi_k/\sqrt{i} \label{eq:xpSDE}, \\ -i \dot{\xi}^z_j &=
-\Gamma \xi^+_j + \sum_k O_{jk} \phi_k /\sqrt{i} \label{eq:xzSDE},
\\ - i \dot{\xi}^-_j &= \frac{\Gamma}{2}\exp{\xi^z_j},
\end{align}
\end{subequations}
where $O_{jk}$ is defined by $\sum_{kl} O_{ki}\mathcal{J}^{-1}_{kl}
O_{lj} = 2 \delta_{ij}$, and we take a symmetrized interaction matrix
$\mathcal{J}_{ij} = \frac{J}{2} (\delta_{i j+1} +\delta_{ i
  j-1})$ \footnote{For system sizes $N$ that are multiples of $4$, we add a
  constant diagonal shift to the interaction matrix $\mathcal{J}$ in
  order to make it diagonalizable; see Appendix \ref{app:diag}}.
Before embarking on a detailed examination of (\ref{eq:isingSDEs}), it is instructive to consider some limiting
cases.  In the non-interacting limit $J=0$, one has $O_{jk}=0$, and (\ref{eq:isingSDEs}) reduces to a set of deterministic equations
which can be solved exactly. As expected, these describe a set of decoupled spins
precessing in an external magnetic field $\Gamma$; see Appendix
\ref{app:isingSDEs}. In the limit $\Gamma=0$, the model
(\ref{eq:isingH}) is purely classical. In this case $\xi^{\pm}_j(t)=0$
for all $t$, while $\xi^z_j(t)$ undergoes exactly solvable Brownian
motion; see Appendix \ref{app:isingSDEs}. For generic values
of $\Gamma$ and $J$, the SDEs~(\ref{eq:isingSDEs})
can be solved numerically, as we highlighted in our previous work
\cite{stochasticApproach}.

Throughout this manuscript, we solve the SDEs using the Euler scheme
\cite{kloeden}. We also set $J=1$ and use a discrete time-step $\Delta
t=10^{-5}$ in all of the figures.  For any non-zero $\Delta t$,
numerical solution algorithms for non-linear SDEs can give rise to
diverging trajectories where the stochastic variables grow without
bound \cite{kloeden,Hutzenthaler2011}; for the Ising SDEs, this effect
is most pronounced for large transverse fields
$\Gamma$. 
Empirically, trajectories are found to monotonically
  grow to numerical infinity when 
\begin{equation}\label{eq:divergence}  
|\dot{\xi}^+_i(t)| \Delta t \geq |\xi^+_i(t)|,
\end{equation}  
i.e. when the increment in $|\xi^+_i(t)|$ in a given time-step exceeds the value of $|\xi^+_i(t)|$. In the case of the Ising model, the increment is given by Eq.~(\ref{eq:xpSDE}). Since the fields $\phi_i$ are of order one and $\Gamma \Delta t$ is typically small, Eq.~(\ref{eq:divergence}) can only be satisfied for large $|\xi^+_i|$. The increment is then dominated by the term proportional to $|\xi^{+2}_i|$, and the requirement~(\ref{eq:divergence}) translates into a divergence condition
  $|\xi^+_i(t)| > \xi^+_c$, where $\xi^+_c \equiv 2/ \Gamma\Delta
  t$. Diverging trajectories can therefore be identified by comparing
  $\xi^+_i(t)$ to $\xi^+_c$ at each time $t$.
With our choice of $\Delta t$, we retain between $99 \%$ and
$100 \%$ of the total number of trajectories, depending on the chosen parameters. 
The stochastic averages are performed by retaining only the non-diverging trajectories at a given time $t$.
Whenever trajectories are excluded, we report their relative fraction in the
associated figure caption. We estimate the magnitude of the fluctuations on our results via the standard error $s_e = {\sigma}/\sqrt{n_B}$, where $\sigma$ is the standard deviation obtained by splitting the data into $n_B=5$
batches of trajectories; we omit the bars when they are
comparable to, or smaller than, the plot points. In order to
illustrate the general approach, we focus on relatively small system
sizes with $N\le 10$ spins. This aids comparison with Exact
Diagonalization (ED) using the QuSpin package \cite{quSpin} and reduces the computational cost, whilst
exposing the main features. We also confine ourselves to times
$t\lesssim 1/J$, before stochastic fluctuations become important. In
Sections \ref{sec:fluctuations} and \ref{sec:cost} we will examine the scaling of the method with increasing $N$ and discuss the eventual breakdown with increasing $t$. 
\subsection{Loschmidt Amplitude}
\label{sec:losch}
\begin{figure}
\includegraphics[width=\linewidth]{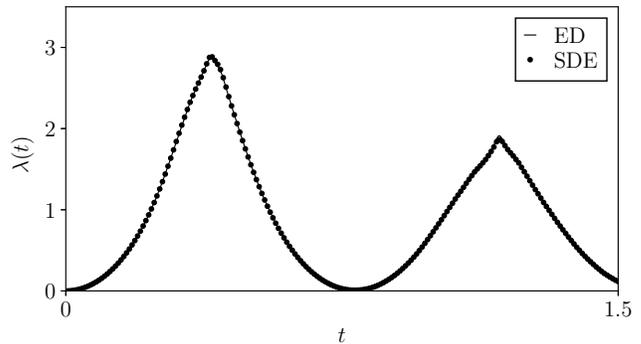}
\caption{Loschmidt rate function $\lambda(t)$ for the 1D quantum Ising model following a quantum quench from the fully-polarized initial state $\ket{\Downarrow}$ to the PM phase with $\Gamma=16\Gamma_c$. The results obtained from the SDEs (filled circles) are in good agreement with ED (solid line) for $N=9$
spins. The SDE results were obtained by averaging over $10^5$
realizations of the stochastic process.  The fraction of diverging trajectories at the stopping time is of order $1\%$.}\label{Fig:loschRfn}
\end{figure}

\begin{figure}
\includegraphics[width=\linewidth]{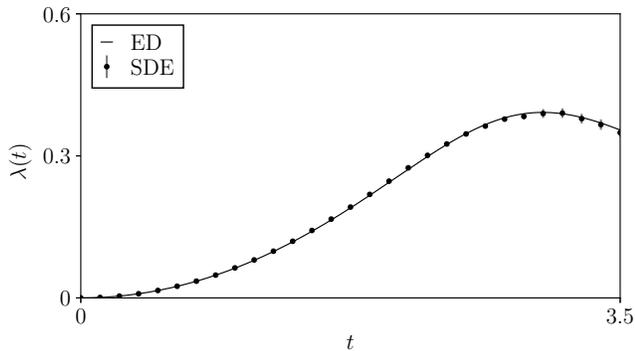}
\caption{Loschmidt rate function $\lambda(t)$ for the 1D quantum Ising
  model following a quantum quench from the initial state
  $\ket{\Rightarrow}$, corresponding to the paramagnetic ground state
  when $\Gamma=\infty$, to the FM phase with $\Gamma=\Gamma_c/4$. The
  results obtained from the SDEs (filled circles) are in good
  agreement with ED (solid line) for $N=5$ spins. The SDE results were
  obtained by averaging over $10^5$ realizations of the stochastic
  process.}\label{Fig:loschFromPM}
\end{figure}

In order to illustrate the general approach, we begin by considering
the Loschmidt amplitude for different quantum quenches. For systems
initialized in the fully-polarized state $ \ket{\Downarrow} \equiv
\otimes_i \ket{\downarrow}_i $, corresponding to a FM ground
  state of the Hamiltonian~(\ref{eq:isingH}) when $\Gamma=0$, the
general formula~(\ref{eq:loschGen}) reduces to
\begin{equation}\label{eq:survivalAmplitudeStochastic}
A(t) = \Big\langle \prod_i^N \exp\left(-\frac{\xi^z_i(t)}{2}\right) \Big\rangle_\phi,
\end{equation}
as reported in our previous work \cite{stochasticApproach}. In
Fig.~\ref{Fig:loschRfn} we show the time-evolution of $\lambda(t)$
following a quench from $\Gamma=0$ to $\Gamma=16\Gamma_c$, across the
quantum phase transition at $\Gamma_c$. The results obtained from the
numerical solution of the SDEs in (\ref{eq:isingSDEs}) are in good
agreement with ED. They also correctly reproduce the sharp peak in
$\lambda(t)$ corresponding to a DQPT in the thermodynamic limit
\cite{heyl2013}.
Going beyond our previous work \cite{stochasticApproach}, it is also
possible to consider quenches from the PM phase to the FM phase. For
example, for quenches starting in the PM ground state
$\ket{\Rightarrow}\equiv \otimes_i \ket{\rightarrow}_i $ for
$\Gamma=\infty$, where $\hat
S_i^x\ket{\rightarrow}_i=1/2\ket{\rightarrow}_i$, the general formula
(\ref{eq:loschGen}) reduces to
\begin{equation}\label{eq:fromPM}
A(t) = \Big\langle \prod_i^N \frac{1}{2} e^{-\frac{\xi^z_i}{2}} \left(\xi^-_i \xi^+_i+\xi^-_i+\xi^+_i+e^{\xi^z_i}+1\right) \Big\rangle_\phi .
\end{equation}
In Fig.~\ref{Fig:loschFromPM} we show the time-evolution of $\lambda(t)$ following a quench
from $\Gamma=\infty$ (PM) to $\Gamma=\Gamma_c/4$ (FM), computed from Eq.~(\ref{eq:fromPM}). Again, we find very good agreement with ED. 
It is worth noting that, in contrast to the simple result
(\ref{eq:survivalAmplitudeStochastic}), the expression
(\ref{eq:fromPM}) features a sum of terms inside the average. However,
from a computational standpoint, this only involves a linear increase
in the number of operations required. Furthermore, the averaging need
not be performed at each time-step: while for numerical accuracy the
SDEs are solved with a small time-step (e.g. $\Delta t \approx
10^{-5}$), observables may be computed on a coarser time interval
(e.g. $\Delta \bar{t} \approx 10^{-3}$). The main computational cost
of the method is associated with solving the SDEs, rather than
performing the averages. Therefore, the presence of the additional
terms in Eq.~(\ref{eq:fromPM}) does not significantly affect the
computational cost: this applies to all the other examples considered
in this Section. Finally, we note that Eq.~(\ref{eq:fromPM}) can be
evaluated from the same set of trajectories as used in
Eq.~(\ref{eq:survivalAmplitudeStochastic}). Thus, in contrast with
other numerical techniques such as time-dependent Density Matrix
Renormalization Group (tDMRG) approaches or ED, the same data can be
used to compute the time-evolution of different initial states.

\begin{figure}
\centering
\includegraphics[width=\linewidth]{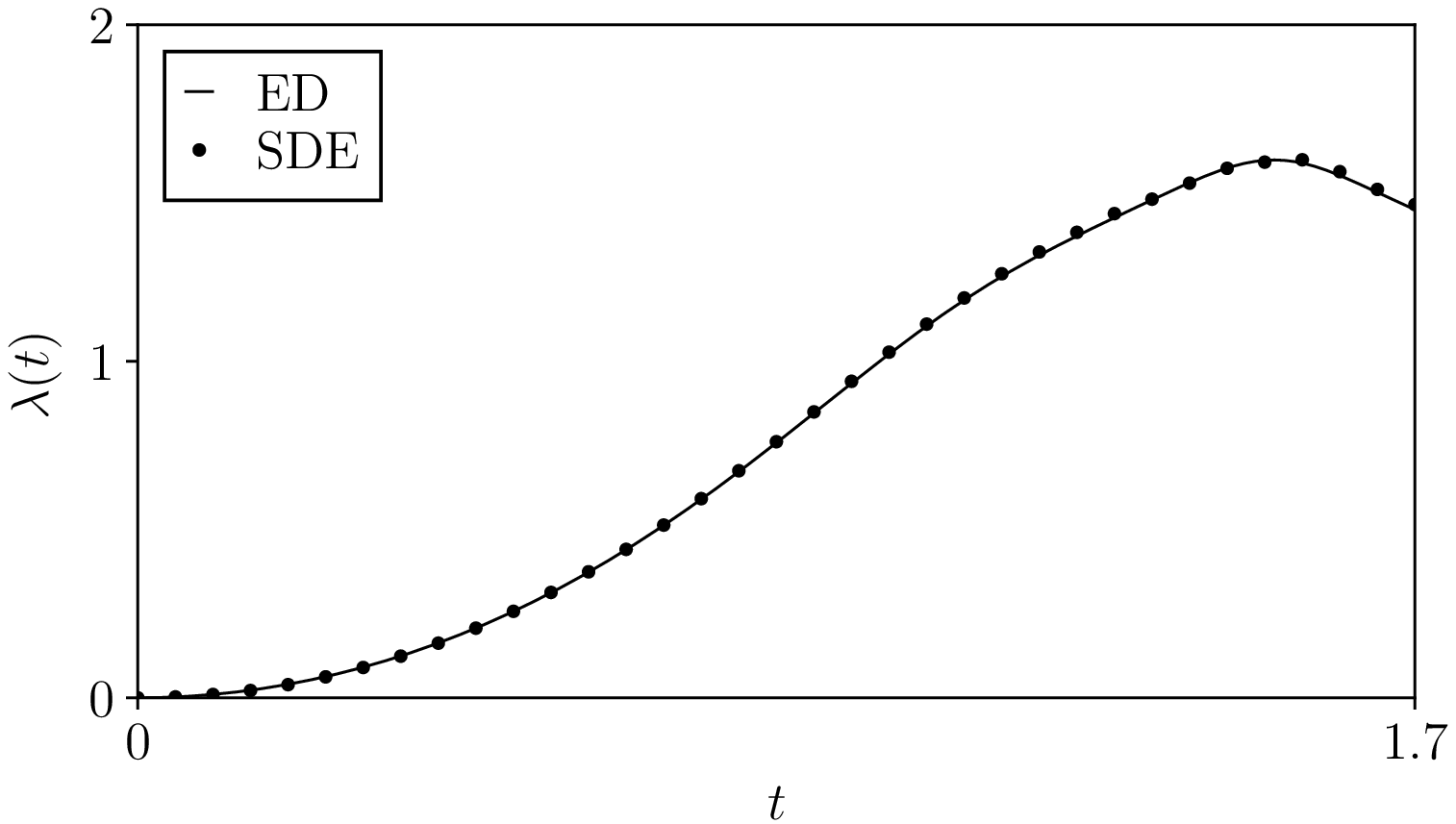}
 \caption{Loschmidt rate function $\lambda(t)$ for the 1D quantum Ising model
following a quantum quench from the spatially inhomogeneous initial
state $\ket{\uparrow \downarrow \downarrow \downarrow \downarrow}$ to
the PM phase with $\Gamma=4 \Gamma_c$. The results obtained from the
SDEs (filled circles) are in good agreement with ED (solid line) for
$N=5$ spins. The SDE results were obtained by averaging over $10^5$
realizations of the stochastic process. Less than $0.1\%$ of the
trajectories were found to be divergent at the stopping time.}\label{Fig:l1S} 
\end{figure}

\begin{figure}
\centering
\includegraphics[width=\linewidth]{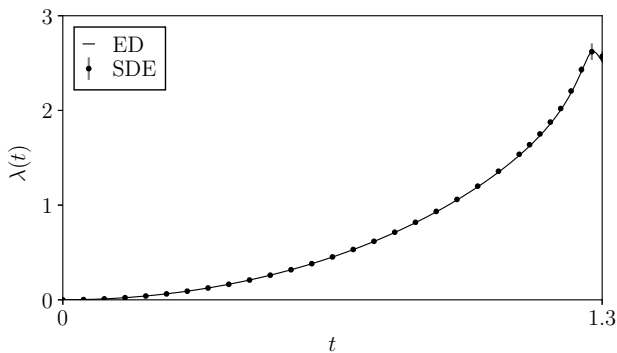}
\caption{Loschmidt rate function $\lambda(t)$ for the 1D quantum Ising model
following a quantum quench from the spatially inhomogeneous initial
state $\ket{\uparrow \uparrow \uparrow \downarrow \downarrow}$ to the
PM phase with $\Gamma=4 \Gamma_c$. The results obtained from the SDEs
(filled circles) are in good agreement with ED (solid line) for $N=5$
spins. The SDE results were obtained by averaging over $10^5$
realizations of the stochastic process. Less than $0.1\%$ of the
trajectories were found to be divergent at the stopping time. Larger
error bars are visible in the vicinity of the peak, due to the
presence of enhanced stochastic fluctuations.}
 \label{Fig:lDW} 
\end{figure}

As discussed in Section \ref{subsec:loschmidt}, the stochastic
approach can also handle spatially inhomogeneous initial states. For
example, we may consider domain wall initial conditions:
\begin{align}
  \ket{\psi(0)} = \ket{\uparrow}_1\otimes  \dots \ket{\uparrow}_{M} \otimes \ket{\downarrow}_{M+1} \dots \otimes \ket{\downarrow}_N,
  \label{eq:domwallinit}
\end{align}
where $1\le M< N$. In this case
\begin{equation}\label{eq:nonUnInState}
A(t) = \Big\langle \prod_{j=1}^{M} \left(e^{\xi_j^z}+\xi^-_j\xi^+_j \right)  \prod_{i=1}^{N} e^{-\frac{\xi^z_i(t)}{2}} \Big\rangle_\phi  .
\end{equation} 
In Figs~\ref{Fig:l1S} and \ref{Fig:lDW} we show the
results for $\lambda(t)$ for different values of $M$, corresponding to a single spin flip and an extended domain of inverted spins
respectively. Once again, the results are in good agreement with
ED.
 
\subsection{Magnetization Dynamics}
\begin{figure}
\includegraphics[width=\linewidth]{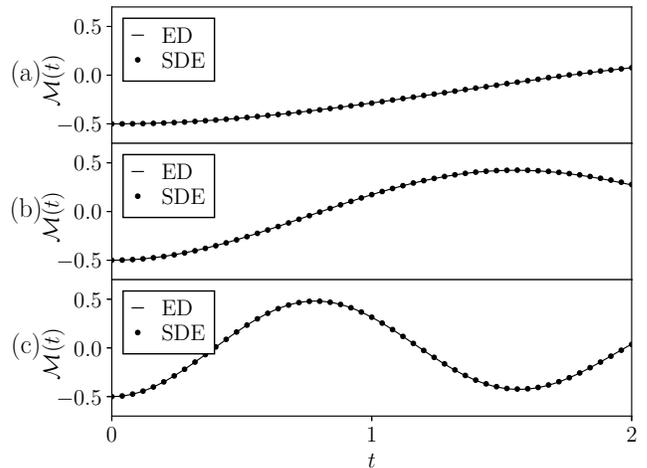}
\caption{Time-evolution of the magnetization $\mathcal{M}(t)$ for the 1D quantum Ising model following quantum quenches from the fully-polarized initial state $\ket{\Downarrow}$ to different values of the final transverse field. (a) $\Gamma=2\Gamma_c$, (b)
$\Gamma=4\Gamma_c$, (c) $\Gamma=8\Gamma_c$. The SDE results (filled circles) computed from (a) $2\times 10^5$ (b) $3\times 10^5$ and (c) $4\times 10^5$ trajectories, are in good agreement with ED (solid line) for $N=3$ spins.}\label{Fig:magnetization} 
\end{figure}

A key observable for non-equilibrium quantum spin systems is the
time-dependent magnetization $\mathcal{M}(t) = \sum_i^N
\mathcal{M}_i/N$ where $\mathcal{M}_i(t) = \langle \hat{S}^z_i(t)
\rangle$. Here we consider quantum quenches from the initial state
$\ket{\Downarrow}$ to different final values of $\Gamma$ in the PM
phase. As can be seen in Fig.~\ref{Fig:magnetization}, the results
obtained by performing the stochastic average in
(\ref{eq:generalMagnetization}) are in very good agreement with ED;
here we focus on small system sizes with $N=3$ spins as we need to
average over two sets of disentangling variables, $\xi_i$ and
$\tilde\xi_i$.
Again, we may consider different initial conditions, such as the
inhomogeneous state (\ref{eq:domwallinit}).
For example, for an initial state where the spin at site $i$ is
pointing up and every other spin is pointing down, the time-dependent
magnetization at site $i$ is given by
\begin{widetext}
\begin{equation}\label{eq:magdomain}
{\mathcal M}_i(t)=  \langle \hat{S}^z_i(t) \rangle = -\frac{1}{2}  \Big\langle e^{-\sum_j  \left( \frac{\xi^z_j+\tilde{\xi}^{z*}_j}{2} \right) } \left[ (e^{\xi^z_i} + \xi^-_i \xi^+_i) (e^{\tilde{\xi}^z_i }+ \tilde{\xi}^-_i \tilde{\xi}^+_i)^*) - \xi^-_i \tilde{\xi}^{-*}_i \right]  \prod_{j\neq i}(1+\xi^+_j \tilde{\xi}^{+*}_j)  \Big\rangle_{\phi,\tilde{\phi}},
\end{equation}  
\end{widetext}
where the disentangling variables $\xi$ satisfy the Ising
SDEs~(\ref{eq:isingSDEs}). This result can be obtained using the
  building blocks given in (\ref{eq:normalizations}) and (\ref{eq:szs}) of Appendix \ref{app:buildingBlocks}.

\begin{figure}
\includegraphics[width=\linewidth]{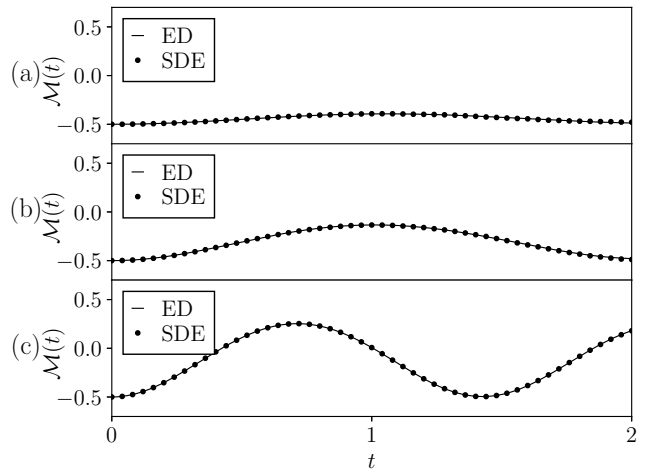}
 \caption{Time-evolution of $\mathcal{M}(t)$ for the 1D quantum Ising
model with an integrability-breaking longitudinal field $h=2J$. We
consider quantum quenches from the fully-polarized initial state
$\ket{\Downarrow}$ to (a) $\Gamma=J$, (b) $\Gamma=2J$, and (c)
$\Gamma=4J$. The SDE results (filled circles) computed from (a) $5\times 10^5$ (b) $2\times 10^5$ and (c) $6\times 10^5$
trajectories are in good agreement with ED (solid line) for $N=3$
spins.}\label{Fig:magnetizationNint} 
 \end{figure}

A significant feature of the stochastic approach to non-equilibrium
quantum spin systems is that it is not restricted to integrable
models. A simple way to break the integrability of the quantum Ising
model (\ref{eq:isingH}) is through the addition of a longitudinal
magnetic field $h$, so that the Hamiltonian is given by $\hat{H} =
\hat{H}_I + h \sum_j \hat{S}^z_j$. In the stochastic formalism the
dynamics of this non-integrable model is described by the Ito SDEs
\begin{subequations}
\begin{align}\label{eq:nonIntIsingSDEs}
-i \dot{\xi}^+_j &= \frac{\Gamma}{2} (1-{\xi^+_j}^2) - h \xi^+_j + \xi^+_j  \sum_k O_{jk} \phi_k/\sqrt{i} ,  \\
-i \dot{\xi}^z_j &= -h -\Gamma \xi^+_j  + \sum_k O_{jk} \phi_k /\sqrt{i} , \\
-i \dot{\xi}^-_j &= \frac{\Gamma}{2}\exp{\xi^z_j},
\end{align}
\end{subequations}
where $O$ is the same as for the purely transverse field Ising model, as given in Section \ref{sec:ising}. In Fig.~\ref{Fig:magnetizationNint}, we show
results for ${\mathcal M}(t)$ corresponding to quenches from the fully-polarized initial state $\ket{\Downarrow}$ to different values of
$\Gamma$, with $h$ held fixed. Once again, we find very good agreement
with ED. It is interesting to note that the same formula
(\ref{eq:generalMagnetization}) governs the dynamics in both the
integrable and non-integrable cases; the Hamiltonian enters only via
the time-evolution of the disentangling variables $\xi^{a}_i$, not the function being averaged.

\begin{figure}
\includegraphics[width=\linewidth]{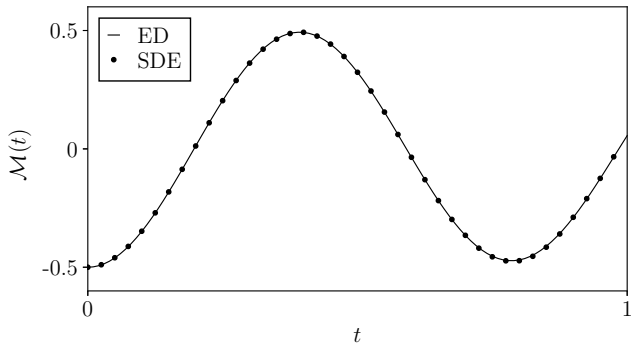}
\caption{Time-evolution of $\mathcal{M}(t)$ for a $2\times 3$ site 
  quantum Ising model following a quantum quench from the fully-polarized
  initial state $\ket{\Downarrow}$ to $\Gamma=8J$. The SDE results
  (filled circles) obtained from $5\times 10^5$ trajectories, are in
  very good agreement with ED (solid line). Less than $1\%$ of the
  trajectories were found to be divergent at the stopping time.}
\label{fig:magn2D}
\end{figure}

As discussed in our previous work \cite{stochasticApproach}, the
stochastic approach can also be used in higher dimensions. For
simplicity, we focus on the two-dimensional (2D) quantum Ising model
with the Hamiltonian
\begin{equation}
  \hat H_{\rm I}^{\mathrm{2D}}= -J\sum_{\langle \mathbf{i} \mathbf{j} \rangle} \hat S_{\mathbf{i}}^z\hat S_{\mathbf{j}}^z - \Gamma \sum_{\mathbf{i}}
  \hat S_{\mathbf{i}}^x,
  \label{Ising2D}
\end{equation}
where $\mathbf{i}$ and $\mathbf {j}$ indicate sites on a square
lattice.  In equilibrium, this model exhibits a quantum phase
transition when $\Gamma=\Gamma_c^{2D}\approx 1.523J$ \cite{pfeuty1971,DuCroodeJongh1998}. In
Fig.~\ref{fig:magn2D} we show results for the magnetization dynamics
${\mathcal M}(t)$ following a quantum quench from the fully-polarized
initial state $\ket{\Downarrow}$ to $\Gamma\approx 5.3 \Gamma_c^{2D}$. Again,
the results are in very good agreement with ED. This highlights that the
stochastic formula for $\langle\hat S_i^z(t)\rangle$ in higher
dimensions is the immediate generalization of the 1D result
(\ref{eq:generalMagnetization}), where the products are extended over all the lattice
sites. The same holds true for other local observables.

\section{Dynamics of the disentangling variables}\label{sec:classical}
A notable feature of the stochastic approach to 
quantum spin systems is that it allows the derivation of exact
stochastic formulae such as equations (\ref{eq:loschGen}) and (\ref{eq:generalMagnetization}). In this
framework, time-dependent quantum expectation values are obtained by
averaging explicit functions of the classical stochastic variables, 
$\xi^a_i(t)$. It is therefore interesting to investigate to what
extent the quantum dynamics is reflected in these classical variables.
\subsection{Distributions of the Classical Variables}
\begin{figure}
\includegraphics[width=\linewidth]{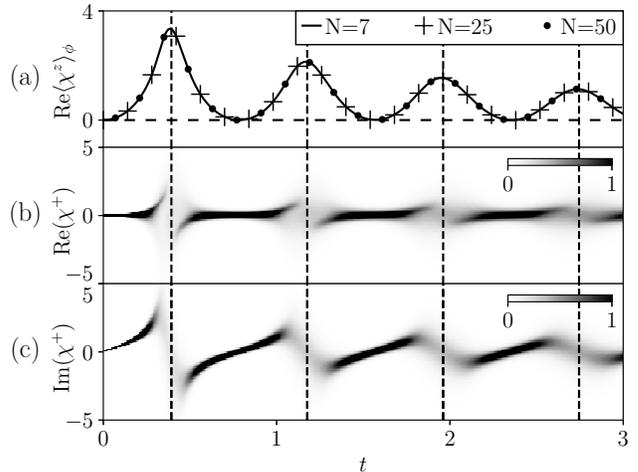}
 \caption{Time-evolution of the disentangling variables in the 1D
   quantum Ising model following a quantum quench from the
   fully-polarized initial state $\ket{\Downarrow}$ to $\Gamma=16
   \Gamma_c$. (a) Dynamics of Re\,$\langle \chi^z(t)\rangle_\phi$ for $N=7, 25, 50$ spins showing maxima in the vicinity of the Loschmidt
   peak times. The latter are obtained by ED (dashed lines). Dynamics
   of the distribution of (b) ${\rm Re}\,\chi^+(t)$ and (c) ${\rm
     Im}\,\chi^+(t)$ for $N=7$ spins showing signatures of the
   DQPTs.}
  \label{Fig:chiP} 
 \end{figure}

\begin{figure}
\includegraphics[width=\linewidth]{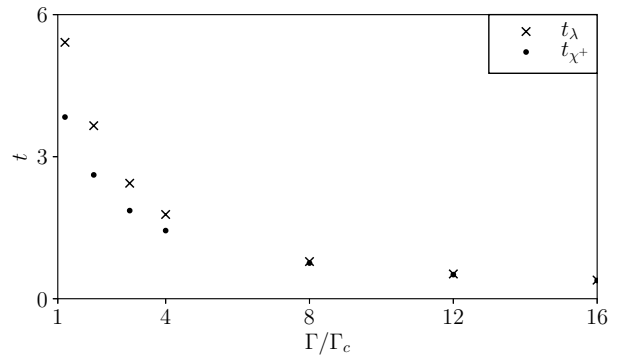}
\caption{Comparison of the time $t_\lambda$ of the first Loschmidt
  peak obtained by ED (crosses) and the time $t_{\chi^+}$ of the zeros
  of ${\rm Im}\langle \chi^+(t)\rangle_\phi$ (dots) for the 1D quantum
  Ising model with $N=7$ sites. The data correspond to quantum
  quenches from the fully-polarized initial state $\ket{\Downarrow}$
  to different values of $\Gamma$. In the limit of large $\Gamma$ the
  results coincide, but for small $\Gamma$, the results of ED differ
  from those given by the approximation ${\rm Im}\langle
  \chi^+(t)\rangle_\phi=0$. Note that for $\Gamma <\Gamma_c$ (not
  shown), there are no DQPTs for $N\rightarrow\infty$. However, zeros
  of ${\rm Im}\langle \chi^+(t)\rangle_\phi$ persist for
  $\Gamma<\Gamma_c$; these get pushed to later times as $\Gamma$ decreases. }
\label{Fig:quenchParameter} 
\end{figure}

As we discussed in Section \ref{sec:losch}, the stochastic formula for
the Loschmidt amplitude has a particularly simple form for quantum
quenches starting in the fully-polarized initial state
$\ket{\Downarrow}$. In this case $A(t)$ can
be written as \cite{stochasticApproach}
\begin{equation}\label{eq:loschFromClassical}
A(t)= \langle e^{-\frac{N}{2} \chi^z(t)} \rangle_\phi,
\end{equation}
where we define the site-averaged variables $\chi^a(t)\equiv N^{-1}
\sum_i \xi^a_i(t)$. It is readily seen that the Loschmidt amplitude is
directly determined by the statistical properties of $\chi^z(t)$. In
particular, the functional form of (\ref{eq:loschFromClassical})
suggests that the peaks in $\lambda(t)\equiv -N^{-1}\log|A(t)|^2$
occur in close proximity to (although not necessarily coincident with) the peaks in the distribution of
$\chi^z(t)$, and its classical average $\langle \chi^z(t)\rangle_\phi$ \cite{stochasticApproach}. In Fig.~\ref{Fig:chiP}(a)
we show the time-evolution
of the latter, which indeed exhibits maxima in the vicinity of the Loschmidt
peaks, and has little dependence on system size.  In addition, the
turning points of $\langle \chi^z(t)\rangle_\phi$ coincide with the
zeros of $\langle \chi^+(t)\rangle_\phi$ due to the exact relation
$i\langle\dot\chi^z(t)\rangle_\phi=\Gamma\langle\chi^+(t)\rangle_\phi$,
which follows from the Ising SDE in Eq.~(\ref{eq:xzSDE}) \cite{stochasticApproach}.  In
general, it is important to stress that the average of
the exponential in (\ref{eq:loschFromClassical}) is {\em not} the
exponential of the average, $-N\langle
\chi^z(t)\rangle_\phi/2$.
As such, the turning points of $\langle \chi^z(t)\rangle_\phi$ are not in general located at the exact positions of the Loschmidt peaks. In
Fig.~\ref{Fig:quenchParameter}
we show the comparison between the turning points of $\langle
\chi^z(t)\rangle_\phi$, or equivalently the zeros of $\langle
\chi^+(t)\rangle_\phi$, and the Loschmidt peak times obtained via ED,
for different quantum quenches. It is evident that these quantities
are in excellent quantitative agreement for quenches with $\Gamma\gg
\Gamma_c$, but differ for $\Gamma\sim \Gamma_c$. This can be
understood from the Ising SDEs in
(\ref{eq:isingSDEs}). In the limit $\Gamma\rightarrow \infty$ the
equations become deterministic and the average of the exponential in
(\ref{eq:loschFromClassical}) is equal to the exponential of the
average; away from this limit, this is not the case. Nonetheless, the
exact formula (\ref{eq:loschFromClassical}) still applies, and its
predictions are in good agreement with ED.

\begin{figure}
\includegraphics[width=\linewidth]{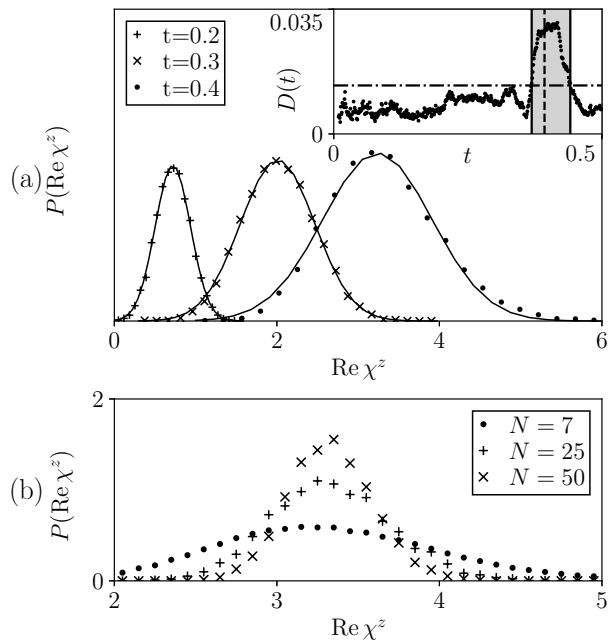}
 \caption{(a) Time-evolution of the distribution of ${\rm
     Re}\,\chi^z(t)$ for the 1D quantum Ising model with $N=7$
   spins. The data correspond to a quantum quench from the
   fully-polarized initial state $\ket{\Downarrow}$ with $\Gamma=0$ to
   $\Gamma=16\Gamma_c$, at times $t=0.2$ (plus signs), $t=0.3$
   (crosses) and $t=0.4$ (dots). We employ different normalizations
   for $P(\text{Re}\,\chi^z(t))$ at different $t$ for ease of
   visualization. The distribution broadens on approaching the
   Loschmidt peak at $t=0.39$, and narrows afterwards. The
   distribution is approximately Gaussian away from the Loschmidt
   peaks, as indicated by the Gaussian fits (solid lines), but is
   non-Gaussian in their vicinity. Inset: Time-evolution of the KS
   statistic $D(t)$ on passing through the first Loschmidt peak. We
   compare the value of $D(t)$ to the critical value of $D_c(\alpha)$
   corresponding to the chosen significance of $\alpha=5 \%$
   (dashed-dotted line). When $D>D_c(\alpha)$, the distribution can be regarded as non-Gaussian. This is observed in the shaded region near the DQPT, whose position is indicated by the dashed vertical line. (b) Variation of the distribution of $\text{Re}\,\chi^z(t)$ at $t=0.39$ with increasing system size $N$. The distribution becomes more sharply peaked as $N$ increases.}\label{Fig:chiZprofile}
\end{figure}

Signatures of the DQPTs can also be seen in the distributions of the
classical variables, which show marked features and enhanced
broadening in their vicinity as shown in Figs~\ref{Fig:chiP}(b) and
(c). In particular, the distribution of ${\rm Re}\,\chi^z(t)$ is
approximately Gaussian away from the DQPTs, but is non-Gaussian in
their proximity, as illustrated in Fig.~\ref{Fig:chiZprofile}(a).

The departures from Gaussianity can be quantified by using the
Kolmogorov--Smirnov (KS) test \cite{James2006}. In this test, one
  considers the KS statistic $D$, which measures the deviation of the
  observed distribution $P\left(x\right)$ of a variable $x$ from the
  best-fitting Gaussian distribution $P_G \left(x\right)$:
\begin{align}\label{eq:kolmogorovSmirnov}
D \equiv \max_{x} \vert P \left(x \right) - P_G\left(x \right)\vert.
\end{align} 
The aim of the test is to accept or reject the null hypothesis that
the observed data come from a Gaussian distribution, to a given
statistical significance. The statistical significance $\alpha$ is
defined as the probability that the test rejects the null hypothesis
when this is in fact true, i.e. the probability that the test fails to
recognize a Gaussian distribution. The statistical significance
$\alpha$ determines a critical value $D_c(\alpha)$ for which the null
hypothesis is rejected with significance $\alpha$ when
$D>D_c(\alpha)$.  For a sufficiently large number of samples
$\mathcal{N}$, the limiting distribution of $D$ is given by
\cite{James2006}
\begin{align}
P(\sqrt{\mathcal{N}} D > z) = 2 \sum_{r=1}^\infty (-1)^{r-1} e^{-2 r^2 z^2}.
\end{align}
The critical value $D_c(\alpha)$, for a given $\alpha$ and number of samples $\mathcal{N}$, is then given by $D_c(\alpha) = z_c(\alpha)/\sqrt{\mathcal{N}}$, where $z_c(\alpha)$ is determined by solving $P(\sqrt{\mathcal{N}} D>z_c)=\alpha$.

To analyze the distribution of $\text{Re}\,\chi^z(t)$, we evaluate (\ref{eq:kolmogorovSmirnov}) with $x=\text{Re}\,\chi^z(t)$. The results of the KS test are shown in the inset of Fig.~\ref{Fig:chiZprofile}(a). The null hypothesis is rejected at the $\alpha=5 \%$ significance level in the
shaded region surrounding the DQPT, indicating that the distribution of $\text{Re}\,\chi^z(t)$ is non-Gaussian in this region. This behavior persists for
different system sizes, with the distributions becoming narrower as
$N$ increases, as shown in Fig.~\ref{Fig:chiZprofile}(b).

\subsection{Bounds}
\begin{figure}
\includegraphics[width=\linewidth]{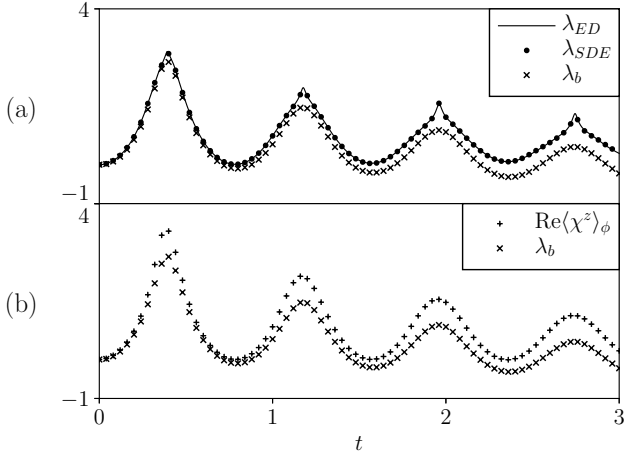}
 \caption{(a) Time-dependent lower bound $\lambda_b(t)$ (crosses) on
   the Loschmidt rate function following a quantum quench in the 1D
   quantum Ising model from $\Gamma=0$ to $\Gamma=16\Gamma_c$. The
   data correspond to the solution of the Ising SDEs (filled circles),
   ED (solid line) and Eq.~(\ref{eq:lambdab}) with $f(\chi(t))=e^{-N\chi^z(t)/2}$ (crosses) for an
   $N=7$ site system. (b) The approximation to the Loschmidt rate
   function ${\rm Re}\langle\chi^z(t)\rangle_\phi$ is also bounded by
   $\lambda_b(t)$.  For large values of $\Gamma$, $\lambda(t)$, ${\rm
     Re}\langle \chi^z(t)\rangle_\phi$ and $\lambda_b(t)$
   coincide.}
 \label{Fig:bound} 
\end{figure}

The stochastic approach also enables one to derive bounds on the
Loschmidt rate function,
$\lambda(t)\equiv -N^{-1}\ln|A(t)|^2$, where $A(t)=\langle
f(\chi(t))\rangle_\phi$ and the function $f(\chi(t))$
depends on the initial conditions. Using the fact that $|\langle f(\chi(t))\rangle_\phi|\le \langle |f(\chi(t))|\rangle_\phi$ one immediately obtains
$\lambda(t)\ge \lambda_b(t)$ where 
\begin{equation}
  \lambda_b(t)\equiv -\frac{2}{N}\ln \langle |f(\chi(t))|\rangle_\phi.
  \label{eq:lambdab}
  \end{equation}
This is confirmed in Fig.~\ref{Fig:bound}(a), where we consider
quenches from the fully-polarized initial state $\ket{\Downarrow}$,
corresponding to $f(\chi(t))=e^{-N\chi^z(t)/2}$. Application of
Jensen's inequality \cite{inequalities} in this case also shows that
\begin{equation}\label{eq:bound}
  {\rm Re}\langle \chi^z(t)\rangle_\phi=-\frac{2}{N}\ln|e^{-\frac{N}{2}\langle \chi^z(t) \rangle_\phi}|\ge \lambda_b(t),
\end{equation}
as confirmed in Fig.~\ref{Fig:bound}(b). 
As $\Gamma\rightarrow\infty$, the three quantities $\lambda(t)$,
$\lambda_b(t)$ and ${\rm Re}\langle \chi^z(t)\rangle_\phi$ all approach the non-interacting result, given in Eq.~(\ref{eq:xzNI}) in Appendix \ref{app:isingSDEs}. In this
limit, as mentioned above, the SDEs (\ref{eq:isingSDEs}) become purely deterministic and
it is possible to replace the average of the exponential in
Eq.~(\ref{eq:loschFromClassical}) with the exponential of the
average. As such, $\lambda(t)$ approaches ${\rm
  Re}\langle\chi^z(t)\rangle_\phi$, in conformity with Fig.~\ref{Fig:chiP}(a).

\subsection{Correlations of the Classical Variables}\label{sec:correlationsClassical}
\begin{figure}
\includegraphics[width=\linewidth]{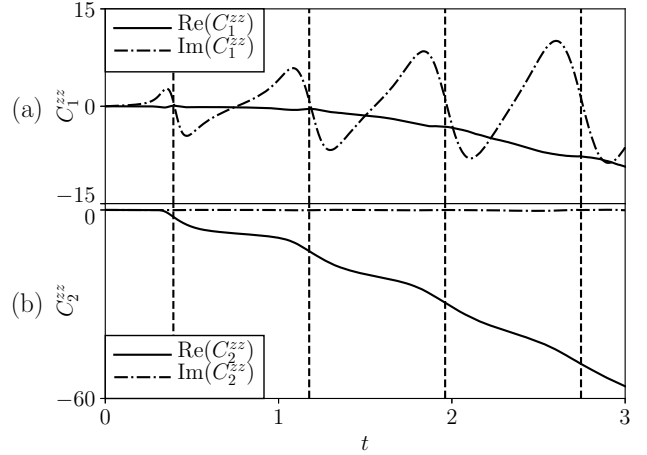}
 \caption{Time-dependent connected correlation functions of the disentangling
variables $\xi^z_i$ following a quantum quench in the 1D quantum Ising
model from $\Gamma=0$ to $\Gamma=16 \Gamma_c$. (a) The first neighbor
correlation function ${\mathcal C}_1^{zz}(t)$ has a monotonically
decreasing real part and an oscillating imaginary part, with zeros
occurring in the vicinity of the Loschmidt peaks (dashed lines,
obtained from ED). (b) The second neighbor correlation function
${\mathcal C}_2^{zz}(t)$ has a vanishing imaginary part, but the real
part decreases monotonically. The latter exhibits steeper gradients in
the vicinity of the Loschmidt peaks.}
 \label{Fig:clCorrelationsXz} 
\end{figure}

\begin{figure}
\includegraphics[width=\linewidth , angle=0]{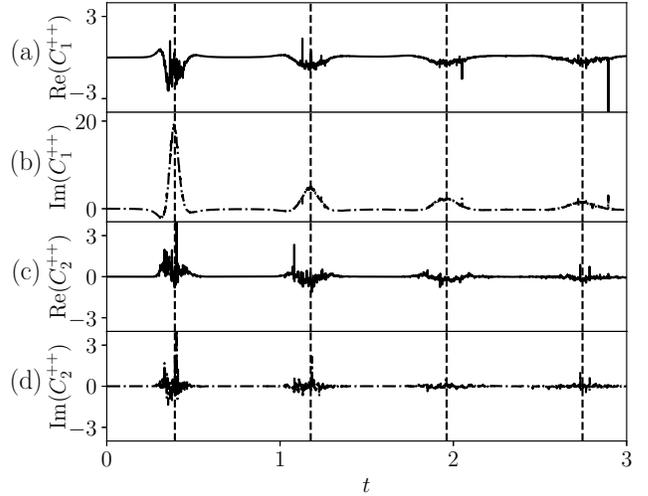}
 \caption{Time-dependent connected correlation functions of the
   disentangling variables $\xi^+_i$ following a quantum quench in the 1D quantum Ising model from $\Gamma=0$ to $\Gamma=16 \Gamma_c$. (a-b)
   The first neighbor correlation functions are small except in the
   vicinity of the Loschmidt peaks (dashed lines), when their
   imaginary part exhibits a sharp peak.  (c-d) The second neighbor
   correlation functions vanish away from the Loschmidt peaks, but
   show enhanced fluctuations as the peak times are
   approached.}
 \label{Fig:clCorrelationsXp} 
\end{figure}

The presence of DQPTs is also reflected in the correlation functions
of the disentangling variables. To see this, it is convenient to
define the site-averaged connected correlation function
\begin{equation}\label{eq:correlationsClassical}
{\mathcal C}^{ab}_n(t) \equiv \frac{1}{N} \sum_{i=1}^N \left( \langle \xi^a_i(t)\xi^b_{i+n}(t) \rangle_\phi  - \langle \xi^a_i(t)\rangle \langle \xi^b_{i+n}(t) \rangle_\phi  \right),
\end{equation}
where $n$ indicates the separation between the two sites. As can be
seen in Fig.~\ref{Fig:clCorrelationsXz}(a), ${\rm Re}\,{\mathcal
  C}_1^{zz}(t)$ decreases smoothly over time, but ${\rm Im}\,{\mathcal
  C}_1^{zz}(t)$ exhibits oscillations within an increasing
envelope. In particular, ${\rm Im}\,{\mathcal C}_1^{zz}(t)$ exhibits
zeros in the vicinity of the Loschmidt peaks.  Likewise, the
second-neighbor correlation function ${\rm Re}\,{\mathcal
  C}_2^{zz}(t)$ decreases rapidly in the vicinity of the DQPTs, while
${\rm Im}\,{\mathcal C}_2^{zz}(t)$ remains zero for all $t$, as shown in Fig.~\ref{Fig:clCorrelationsXz}(b); an analytical proof that 
  ${\rm Im}\,C^{zz}_n=0$ when $n\geq 2$ is provided in
  Appendix~\ref{app:moments}.
Fig.~\ref{Fig:clCorrelationsXp} shows an analogous analysis for
${\mathcal C}^{++}_n(t)$. It can be seen that first neighbor
correlation functions take small values everywhere, except in the
vicinity of the DQPTs where ${\rm Im}\,{\mathcal C}_1^{++}(t)$ peaks;
see Fig.~\ref{Fig:clCorrelationsXp}(a). Likewise, the second neighbor
correlation functions vanish on average for all times, but exhibit
strong fluctuations in the vicinity of the Loschmidt peaks; see
Fig.~\ref{Fig:clCorrelationsXp}(b). 
The further neighbor correlation functions (not shown) are found to
behave similarly to the second neighbor case, due to the nearest
neighbor form of the $\mathcal{J}_{ij}$ matrix under consideration. As
prescribed by Eq.~(\ref{eq:EoMs}), the Ito drift is proportional to
$\sum_k O_{ik} O_{jk} = 2 \mathcal{J}_{ij}$, which is only non-zero
when $j = i \pm1$. As a result, the first neighbor correlation
functions ${\mathcal C}_1^{zz}(t)$ and ${\mathcal C}_1^{++}(t)$ are
qualitatively different from their further neighbor
counterparts. In Appendix~\ref{app:moments}, we provide
  further information on the moments of the disentangling variables,
  explicitly identifying a set of averages which vanish at all times.
\section{Fluctuations}\label{sec:fluctuations}
As we have discussed above, the statistical properties of the
disentangling variables play a central role in the stochastic approach
to quantum spin systems. They provide access to time-dependent quantum
expectation values and exhibit notable signatures in the vicinity of DQPTs. As we will
discuss now, the fluctuations in the disentangling variables also
provide insights into the current limitations of the stochastic
approach. From a numerical perspective, the two main sources of error
arise from the non-zero discretization time-step $\Delta t$, and the
finite number of samples ${\mathcal N}$. The former is relatively
benign for short timescales, but eventually leads to divergences in
the stochastic variables at late times \cite{kloeden, Hutzenthaler2011}. This effect is more pronounced in the presence of large transverse fields $\Gamma$, and can be mitigated by reducing the time-step $\Delta t$. The
latter is more important and arises from performing
stochastic averages over a finite number of samples ${\mathcal
  N}$. For a quantum observable $\langle\hat{\mathcal O}(t)\rangle$
corresponding to a stochastic function $f(\xi(t))$, as defined by (\ref{eq:generalObservable}), the formally exact expression is
approximated by
\begin{align}
\langle \hat{\mathcal{O}}(t) \rangle \approx \frac{1}{{\mathcal
    N}}\sum_{r=1}^{\mathcal N} f_r(t) \equiv S_{\mathcal N}(t),
\end{align} 
where $f_r(t)$ is the value of $f(t)=f(\xi(t))$ for a given
realization $r$ of the stochastic process. In the limit ${\mathcal
  N}\rightarrow\infty$, the central limit theorem implies that the
sample average $S_{\mathcal N}(t)$ is Gaussian distributed, even if
the individual $f_r(t)$ are not, provided that $f(t)$ has finite
variance. The resulting Gaussian distribution has mean $\langle
f(t)\rangle$, and standard deviation $\sigma_{\mathcal
  N}(t)=\sigma(t)/\sqrt{{\mathcal N}}$, where $\sigma(t)$ is the
standard deviation of $f(t)$. The fluctuations in $S_{\mathcal N}(t)$
obtained by sampling the SDEs are therefore proportional to
$\sigma(t)$; the value of $\sigma(t)$ thus determines the number of
simulations required to achieve a given accuracy.

\begin{figure}
\includegraphics[width=\linewidth]{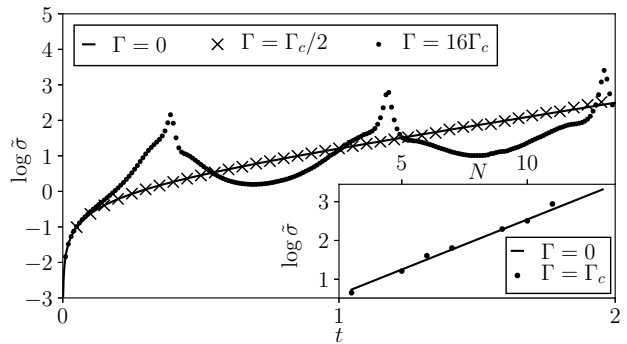}
 \caption{Growth of fluctuations in the stochastic approach. (a)
Time-evolution of the normalized standard deviation
$\tilde{\sigma}(t)$ as defined in the  text, for the Loschmidt
amplitude $A(t)$. We consider quantum quenches in the 1D quantum Ising
model with $N=5$ from the fully-polarized initial state
$\ket{\Downarrow}$ state to different values of $\Gamma$. The solid
line shows the analytical result for the classical case $\Gamma=0$,
corresponding to exponential growth with $t$ and $N$. For
$\Gamma>\Gamma_c$, stronger fluctuations become visible in the
vicinity of the Loschmidt peaks, but the overall growth is consistent
with the classical case. Inset: growth of fluctuations with
increasing $N$, for fixed $\Gamma$ and $t=1$. The results
are consistent with exponential growth.}\label{Fig:fluctuations} 
\end{figure}

\begin{figure}
\centering
\includegraphics[width=\linewidth]{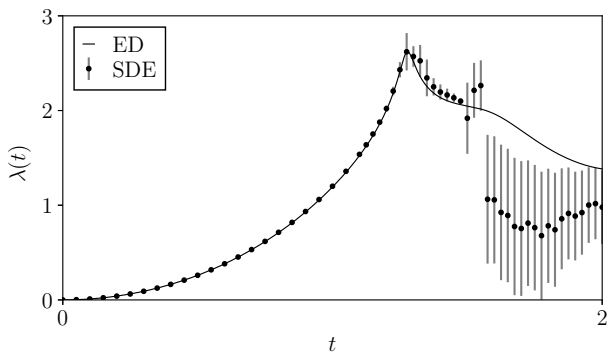}
 \caption{Loschmidt rate function $\lambda(t)$ for the same quantum quench considered in Fig.~\ref{Fig:lDW}, but extending the simulation time. As can be seen from the vertical bars (light grey), for $t\gtrsim 1.3$ strong fluctuations in the disentangling variables hamper the convergence of the stochastic averages (filled dots) to   the results obtained by ED (solid line).}
 \label{Fig:lDW_breakdown} 
 \end{figure}

In order to quantify the growth of fluctuations it is instructive to
consider the Loschmidt amplitude $A(t)$ given by
(\ref{eq:survivalAmplitudeStochastic}), for quenches starting in the
fully-polarized initial state $\ket{\Downarrow}$. Since the Loschmidt
amplitude is exponentially suppressed with increasing system size, it
is convenient to consider the strength of the fluctuations relative to
the mean, using $\tilde{\sigma}(t) = \sigma(t)/|\langle f(t)
\rangle|$. In the classical limit where $\Gamma=0$, one can show that
$|\langle f(t) \rangle|=1$, so that $\tilde\sigma(t)=\sigma(t)$; see
Appendix \ref{app:isingSDEs}. In this case
\begin{equation}\label{eq:expgrowth}
\tilde\sigma(t)= e^{\frac{JNt}{2}}-1,
\end{equation}
which exhibits exponential growth with time $t$ and system size $N$
according to $ \tilde\sigma(t) \sim e^{J Nt/2}$.  In
Fig.~\ref{Fig:fluctuations} we confirm this dependence numerically for
quenches in the 1D quantum Ising model.  A similar exponential growth
of fluctuations is also observed for $\Gamma<\Gamma_c$, as shown in
Fig.~\ref{Fig:fluctuations}. In the regime $\Gamma>\Gamma_c$, enhanced
fluctuations appear in the vicinity of the Loschmidt peaks, but the
overall growth of fluctuations mirrors that in (\ref{eq:expgrowth}).
Similar behavior is also observed for other observables and for
  different initial conditions. In the case of local observables, the
  presence of two time-evolution operators in Eq.~(\ref{eq:onePoint}) translates to an extra factor of $2$ in the exponent, as shown in Appendix \ref{app:isingSDEs} for the magnetization. The exponential
growth of fluctuations for large system sizes and long times
ultimately limits the stochastic approach in its current form. As the
fluctuations in $\tilde\sigma(t)$ increase, an increasing number of
runs is required for the sample mean $S_{\mathcal N}(t)$ to converge to $\langle f(t)\rangle$. This is consistent with our numerical observations, as illustrated in Figs~\ref{Fig:lDW_breakdown} and \ref{fig:effectOfFluctuations}. The simulations typically breakdown at
a characteristic time $t_b\sim 1/NJ$, when the variance of the
spatially summed and time-integrated HS fields is of order unity; this
can also be seen directly from Eq.~(\ref{eq:expgrowth}).

\begin{figure}\label{fig:effectOfFluctuations}
\includegraphics[width=\linewidth]{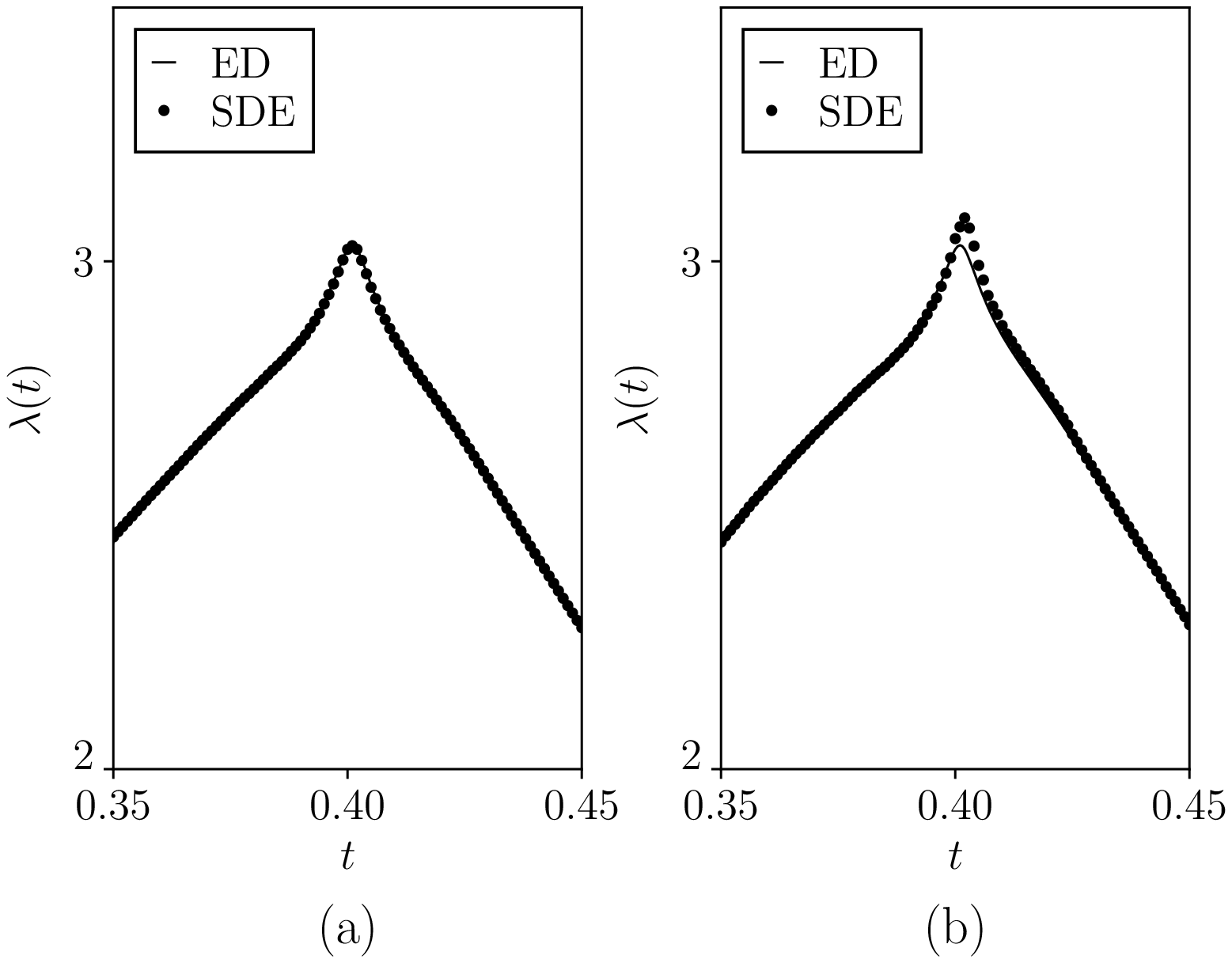}\\
\includegraphics[width=\linewidth]{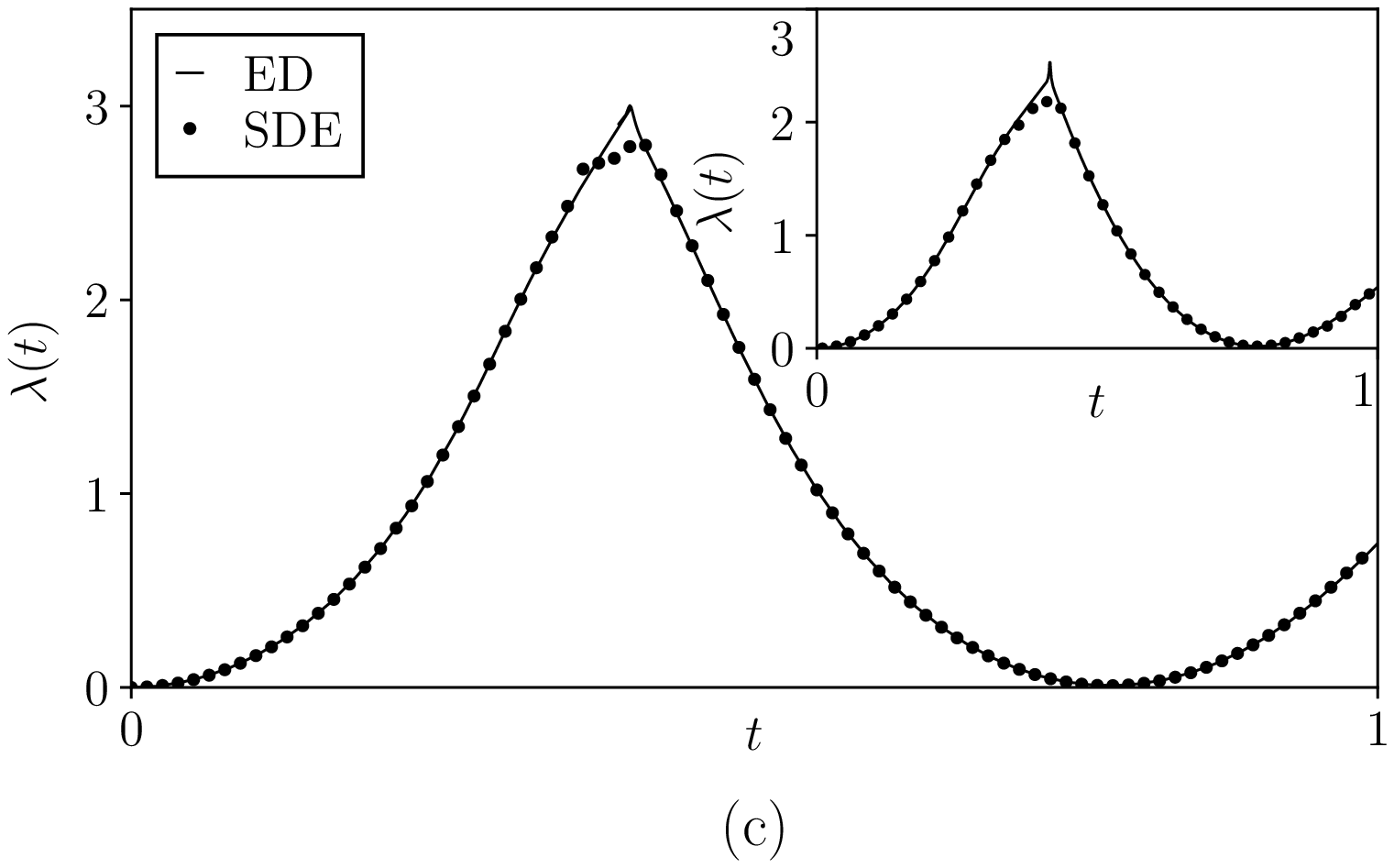}
\caption{(a) Close-up of the first Loschmidt peak for the 1D
  quantum Ising model with $N=14$ spins following a quantum quench
  from $\Gamma=0$ to $\Gamma=16 \Gamma_c$. Using $n=3\times 10^6$
  independent trajectories we reproduce the ED result for
  $\lambda(t)$. (b) Results for the same quench, but with $n=3\times
  10^5$ trajectories. For this smaller number of simulations,
  $\lambda(t)$ converges to the ED result, except in the immediate
  vicinity of the peak. (c) Loschmidt rate function for the quench in
  panel (a), but for $N=21$ spins. For $n=5 \times 10^6$ simulations,
  $\lambda(t)$ at the peak has not yet converged to the ED value. This
  is due to the enhanced fluctuations in the disentangling variables
  $\xi^z_i$ in the vicinity of the peaks, which grow with $N$; see Fig.~\ref{Fig:fluctuations}. Near the peak, the sampling
  is insufficient to reproduce the ED result. However, in all other
  regions of the plot, including times beyond the peak, the result
  obtained from the SDEs is in good agreement with ED. This highlights
  that the method is formally exact, but that sampling is important in
  order to achieve convergence. Inset: analogous results for a
    $5\times 5$ quantum Ising model, corresponding to the upper limit for comparison with ED. The system was initialized in the
    fully-polarized state $\ket{\Downarrow}$ and time-evolved
    with $\Gamma=8J$. The results for $n=4 \times 10^7$ 
    are similar to those in panel (c):
    the SDE results are in good
    agreement with ED before and after the peak, but the sampling is insufficient to resolve the peak.}
\label{fig:effectOfFluctuations}
\end{figure}

\section{Computational Cost}\label{sec:cost}
A notable feature of the stochastic approach to quantum spin systems
is that the numerical solution of the SDEs is intrinsically
parallelizable; the stochastic averages are performed over independent
trajectories and the number of stochastic variables scales linearly in
$N$, due to the HS decoupling of the interactions. The
simulation time also scales linearly with $t$ and ${\mathcal N}$, and
inversely with $\Delta t$. However, as $t$ and $N$ increase, the
exponential growth in the fluctuations requires increasing ${\mathcal
  N}$; this necessitates much longer simulation times than suggested
by the na\"ive linear scaling. Eventually, the averages obtained from
a given number of trajectories fail to converge to the required
quantum expectation values, due to the increasing variance of
the stochastic variables; see Figs~\ref{Fig:fluctuations},
\ref{Fig:lDW_breakdown} and \ref{fig:effectOfFluctuations}.  In the
case of the Loschmidt amplitude, each batch of ${\mathcal N}=10^5$
simulations with $\Delta t=10^{-5}$ takes approximately 1 hour on $96$
cores, per unit interval of time, and per lattice site, i.e. the data
in Fig.~\ref{Fig:loschRfn} correspond to approximately $14$ hours of
simulation time. Local expectation values take a factor of two longer
due to the presence of two sets of disentangling variables. From a
numerical perspective, this is clearly inferior to ED for small system
sizes. However, for larger system sizes, the stochastic approach may
offer some advantages as the number of stochastic variables scales
linearly in $N$. In contrast to ED, one also avoids having to store an
exponentially large matrix in memory. However, this advantage is
offset to some extent as the breakdown time $t_b\sim 1/NJ$ decreases
with increasing system size due to the growth of stochastic
  fluctuations, i.e. there is a trade off between increasing the
system size $N$ and the timescale that can be addressed. We also
  observe slower convergence in the regions where the fluctuations are
  strongest; see Figs~\ref{Fig:fluctuations} and
  \ref{fig:effectOfFluctuations}. This could perhaps be mitigated
through the use of enhanced sampling techniques. Nonetheless, in spite
of these numerical and computational challenges, the stochastic
approach offers a new set of tools for describing non-equilibrium
quantum spin systems. This includes exact stochastic formulae with
wide applicability, which hold in arbitrary dimensions and in the
absence of integrability. The stochastic approach also provides direct
links between quantum and classical dynamics, enabling the transfer of
ideas between different domains of non-equilibrium science.
 
\section{Conclusion}\label{sec:conclusion}
In this work we have investigated a stochastic approach to
non-equilibrium quantum spin systems based on an exact mapping of
quantum dynamics to classical SDEs. We have provided exact stochastic
formulae for a variety of quantum observables, with broad
applicability for spin-$1/2$ systems. We have also outlined the
general approach to express other observables in this framework. We
have illustrated the method in the context of the one- and
two-dimensional quantum Ising model, highlighting the role of the
classical stochastic variables and their relation to dynamical quantum
phase transitions. We have also explored the growth of fluctuations in
the stochastic approach, discussing their scaling with time and
system size, including details of the numerical aspects of the current
implementation of the method. There are many directions for future
research, including the development of improved sampling methods as
well as further exploration of the correspondence between the quantum
and classical dynamics.

We acknowledge helpful conversations with Samuel Begg, John
  Chalker, Andrew Green, Vladimir Gritsev, Lev Kantorovich and Austen
  Lamacraft. MJB is very grateful to John Chalker for early
  discussions on the Hubbard--Stratonovich and stochastic approaches
  to quantum dynamics. 
BD is a Royal Society Leverhulme Trust Senior Research Fellow, ref. SRF\textbackslash R1\textbackslash 180103.   
  SDN acknowledges funding from the Institute of
  Science and Technology (IST) Austria, and from the European Union's
  Horizon 2020 research and innovation programme under the Marie
  Sk\l{}odowska-Curie Grant Agreement No. 754411. SDN also
  acknowledges funding from the EPSRC Centre for Doctoral Training in Cross-Disciplinary Approaches to Non-Equilibrium Systems (CANES) under grant EP/L015854/1. 
  MJB, BD and SDN thank the Centre for
  Non-Equilibrium Science (CNES) and the Thomas Young Centre (TYC). We
  are grateful to the UK Materials and Molecular Modelling Hub for
  computational resources, which is partially funded by EPSRC
  (EP/P020194/1). We acknowledge computer time on the Rosalind High Performance Computer Cluster.

\appendix 
\section{Hubbard--Stratonovich Decoupling of the Time-Evolution Operator}
\label{app:decoupling}
In Section \ref{subsec:HST} we gave a brief overview of the
Hubbard--Stratonovich decoupling of the time-evolution operator
$\hat{U}(t) = \mathbb{T} e^{-i \int_0^t \hat{H}\dd t^\prime}$, where
\begin{align}
\hat{H} &= - \sum_{ijab} \mathcal{J}^{ab}_{ij} \hat{S}^a_i \hat{S}^b_j - \sum_{ia} h^a_i \hat{S}^a_i, 
\end{align}
is  a generic Heisenberg model. Here we provide some of the technical details involved in this procedure. 
Trotter-slicing the time-ordered exponential in $\hat{U}(t)$ one obtains
\begin{widetext}
\begin{align}
\hat{U}(t)&= \mathbb{T} \exp  \left(-i \int_0^t   \mathrm{d}t^\prime  \hat{H}(t^\prime)  \right) = \mathbb{T} \lim_{n\rightarrow \infty} \prod\limits_{m=1}^n \mathrm{exp}\Big(i \Delta t \sum_{ijab} \mathcal{J}_{ij}^{ab}(m \Delta t) \hat{S}_i^a \hat{S}_j^b +i \Delta t \sum_{ja} h^a_j(m \Delta t) \hat{S}^a_j   \Big) ,
\label{fullDiscretised} 
\end{align}
\end{widetext}
where $\Delta t \equiv t/n$. Performing the Hubbard--Stratonovich
transformation at each time slice yields
\begin{widetext}
\begin{equation}\label{eq:optrHSrt}
e^{i \Delta t \sum_{ijab} \mathcal{J}^{ab}_{ij} \hat{S}^a_i \hat{S}^b_j } = \mathcal{C}  \int \prod_{ia} ( \mathrm{d}\varphi^a_i) e^{- \frac{1}{4} \Delta t \sum_{ijab}(\mathcal{J}^{-1})_{ij}^{ab}\varphi^a_i \varphi^b_j + \sqrt{i} \Delta t \sum_{ja} \varphi^a_j \hat{S}^a_j },
\end{equation}
\end{widetext}
where $\mathcal{C}$ is a normalization constant and 
$\varphi^a_i$ are complex scalar fields chosen in such a way as to ensure convergence of the integral in (\ref{eq:optrHSrt}); see the discussion following Eq.~(\ref{eq:eachTerm}) below.
In order to show that Eq.~(\ref{eq:optrHSrt}) holds for spin operators $\hat{S}^a_i$, it is convenient to introduce multicomponent indices $\alpha = \{i,a\}$ so that e.g. $\mathcal{J}_{ij}^{ab}\equiv \mathcal{J}_{\alpha\beta}$:
\begin{equation}\label{eq:multiIndices}
e^{i \Delta t \sum_{ijab} \mathcal{J}^{ab}_{ij} \hat{S}^a_i \hat{S}^b_j } \equiv e^{i \Delta t \sum_{\alpha \beta} \mathcal{J}_{\alpha \beta} \hat{S}_\alpha \hat{S}_\beta}  .
\end{equation}
For simplicity, we assume that the matrix $\mathcal{J}_{\alpha \beta}$ is symmetric, as it can always be redefined so that this is true. Then, $\mathcal{J}_{\alpha \beta}$ can be diagonalized as follows.
We define the matrix $Q$ whose columns are the orthonormal eigenvectors $e^{(\alpha)}$ of $\mathcal{J}$, so that $Q_{\alpha \beta} = e^{(\beta)}_\alpha$:
\begin{equation}\label{eq:Qmat}
Q \equiv   \left({\begin{array}{ccc}
   e^{(1)} & \dots & e^{(3N)} \\
   \downarrow & \dots & \downarrow 
  \end{array} } \right).
\end{equation}
This is an orthogonal matrix satisfying $Q Q^T= Q^T Q =
\mathbb{1}$. We also define the diagonal matrix $D \equiv
\mathrm{diag}(\lambda_1, \dots, \lambda_{3N})$, whose
elements are the (real-valued) eigenvalues of $\mathcal{J}$, arranged in the same order as the columns of $Q$, so
that $Q^T \mathcal{J} Q = D$. Using these, Eq.~(\ref{eq:multiIndices}) can be written as
\begin{align}\label{eq:diagonalOperators}
e^{i \Delta t \sum_{\alpha \beta} \mathcal{J}_{\alpha \beta} \hat{S}_\alpha \hat{S}_\beta}  = e^{ i \Delta t \sum_{\alpha} \lambda_\alpha \hat{\mathscr{S}}_\alpha^2 } .
\end{align}
where we have defined the operators $ \hat{\mathscr{S}}_\alpha \equiv
\sum_{\beta} (Q^T)_{\alpha\beta} \hat{S}_\beta $. For example, for the quantum Ising model~(\ref{eq:isingH}) the $\hat{\mathscr{S}}_\alpha$ are
linear combinations of the $\hat{S}^z_i$ operators at different sites.
One can now factorize the infinitesimal exponentials in
Eq.~(\ref{eq:diagonalOperators}) over $\alpha$ using
\begin{align}\label{eq:factorized}
e^{ i \Delta t \sum_{\alpha} \lambda_\alpha \hat{\mathscr{S}}_\alpha^2 } = \prod_\alpha e^{ i \Delta t  \lambda_\alpha \hat{\mathscr{S}}_\alpha^2 }, 
\end{align}
where we neglect terms of order $(\Delta t)^2$ in the exponent.
For each factor in Eq.~(\ref{eq:factorized}), one obtains
\begin{align}\label{eq:eachTerm}
e^{i \Delta t \lambda_\alpha \hat{\mathscr{S}}_\alpha^2 } = C_\alpha \int  \mathrm{d}\bar{\varphi}_\alpha e^{- \frac{1}{4} \Delta t  \lambda_\alpha^{-1}\bar{\varphi}^2_\alpha  + \sqrt{i} \Delta t\bar{\varphi}_\alpha \hat{\mathscr{S}}_\alpha}  
\end{align}
where $C_\alpha $ is a normalization constant. Since $\hat{\mathscr{S}}_\alpha$ commutes with itself,
the above Gaussian equality can be proved by Taylor expansion of the integrand, where the integration over $\bar{\varphi}_\alpha$
is carried out along the real (imaginary) axis for all positive (negative) eigenvalues
$\lambda_\alpha$. Finally, by changing variables using $\bar{\varphi}_\alpha = \sum_{\beta} (Q^T)_{\alpha \beta}
  {\varphi}_\beta$, the operator identity~(\ref{eq:optrHSrt}) is verified.

\section{Diagonalization of the Noise Action}\label{app:diag}
Following the application of the Hubbard--Stratonovich transformation,
we define the \textit{noise action} $S[\varphi]$ as
\begin{equation}\label{eq:noiseActionInitial}
S[\varphi] \equiv \sum_{ijab}\int_0^t  \frac{1}{4}(\mathcal{J}^{-1})_{ij}^{ab}\varphi^a_i(t^\prime) \varphi^b_j(t^\prime) \mathrm{d}t^\prime .
\end{equation}
We want to perform a change of variables $\varphi^a_i=\sum_{jb}O^{ab}_{ij}\phi^b_j$  so that Eq.~(\ref{eq:noiseActionInitial}) can be recast in the  form
\begin{equation}\label{eq:appnoiseActionDiagonal}
S [\phi] \equiv \sum_{ia} \int_0^t \frac{1}{2} \phi^a_i(t^\prime) \phi^a_i(t^\prime) \mathrm{d}t^\prime.
\end{equation}
For a symmetric interaction matrix $\mathcal{J}^{ab}_{ij}$, one can always construct a matrix $O^{ab}_{ij}$ that diagonalizes (\ref{eq:noiseActionInitial}). 
Using the matrices $Q$ and $D$ introduced in Appendix~\ref{app:decoupling}, the matrix
\begin{align}\label{eq:oMat}
O\equiv \sqrt{2} Q
D^{1/2}
\end{align} 
satisfies $ O^T \mathcal{J}^{-1} O /2 = \mathbb{1}$ and can thus be
used to put the noise action in the desired
form~(\ref{eq:appnoiseActionDiagonal}). The matrix $O$ also satisfies
$O O^T= 2\mathcal{J}$, which is a useful relation when converting the
SDEs between the Ito and Stratonovich conventions; see
Appendix~\ref{app:itoStratonovich}. By writing $O$ in terms of
  its real and imaginary parts $O_R$, $O_I$, one also obtains
  $O_R O_R^T - O_I O_I^T = \mathcal{J}$. Due to the
  definition~(\ref{eq:oMat}), where the matrix $Q$ is real valued and
  the entries of the diagonal matrix $D^{1/2}$ are either purely real
  or purely imaginary, $O$ has either purely real or purely imaginary
  columns. This implies that $O_I O_R^T = O_R O_I^T =0$. We
  will use these properties of $O_R$ and $O_I$ in
  Appendix~\ref{app:isingSDEs}.  The definition of the matrix $O$ is
not unique and depends on the specific ordering of the eigenvalues in
(\ref{eq:Qmat}). This construction breaks down if the interaction
matrix has vanishing eigenvalues and cannot be inverted.  This is
relevant to the quantum Ising model~(\ref{eq:isingH}) for example. In
this case, the interaction matrix is given by
\begin{equation}
\mathcal{J}^{ab}_{ij} = \frac{J}{2} \delta_{az}\delta_{bz} (\delta_{ij+1} + \delta_{i j-1}) \equiv \delta_{az}\delta_{bz} \mathcal{J}_{ij}.
\end{equation}
When the system size $N$ is a multiple of $4$, one of the eigenvalues of $\mathcal{J}$ turns out to be zero.
For spin-$1/2$ systems this can be avoided by including a shift
proportional to the identity operator in the Hamiltonian. Using the
fact that $\sum_i( S^z_i)^2 = \frac{N}{4} \mathbb{1}$, this may be
achieved by adding a term $J_s\delta_{ij} \delta_{az}\delta_{bz}$ to
the interaction matrix. For $J_s \neq 1$, $\mathcal{J}$ becomes
invertible. The corresponding time-evolution operator
acquires a constant phase shift, which does not affect the computation
of physical observables. However, with this modification, $\mathcal{J}_{ii} \neq 0$ and thus $(O O^T)_{ii} \neq 0 $. This leads to
a change in the stochastic equations of motion, as discussed in
Appendix \ref{app:itoStratonovich}.

\section{Disentanglement Transformation}\label{app:disentanglement}
As we discussed in Section \ref{subsec:DT}, the stochastic
time-evolution operator $\hat U_j^s(t)$ can be simplified by means of 
a {\em disentanglement transformation} \cite{hoganChalker,galitski,ringelGritsev}:
\begin{align}
\hat{U}_j^s(t) \equiv \mathbb{T} e^{i \int^t_0 \sum_a \Phi^a_j(t) \hat{S}^a_j } = e^{\xi_j^+(t)
  \hat{S}_j^+} e^{ \xi_j^z(t) \hat{S}_j^z} e^{\xi_j^-(t) \hat{S}_j^-},
\label{app:ujs}
\end{align}
where the explicit group parameterization eliminates the time-ordering
operation. In order to obtain the evolution equations satisfied by the
disentangling variables $\xi_j^a$ \cite{ringelGritsev}, one may
differentiate (\ref{app:ujs}) with respect to time.
Right-multiplying the result by $(\hat{U}^s_j)^{-1}$ one obtains
\begin{equation}\label{eq:disenStart}
(\partial_t \hat U^s_j) ( \hat U^s_j)^{-1} =i \sum_a \Phi^a_j \hat{S}^a_j,
\end{equation}
or, equivalently, 
\begin{equation}\label{eq:disentEquiv}
\sum_a \left( \partial_t \xi^a_j \frac{ \partial \hat U^s_j}{\partial \xi^a_j} \right) ( \hat U^s_j)^{-1}  =i \sum_a \Phi^a_j \hat{S}^a_j.
\end{equation}
For this equality to hold, the coefficients of 
$\hat{S}^a_j$ on each side of Eq.~(\ref{eq:disentEquiv}) must be equal. Considering each spin component $a\in \{+,z,- \}$ in turn, it may be verified that
\begin{align}
\left(\frac{\partial  \hat{U}_j^s}{ \partial \xi_j^+} \right)  (\hat{U}_j^s)^{-1} = \hat S_j^+. 
\end{align}
Similarly,
\begin{align}
  \left( \frac{\partial \hat{U}_j^s}{\partial \xi_j^z} \right) (\hat{U}_j^s)^{-1} =
  \hat S_j^z - \xi_j^+ \hat S_j^+.
\end{align}
In deriving this expression we invoke Hadamard's lemma
\begin{equation}\label{eq:hadamard}
e^{\hat{A}} \hat{B} e^{-\hat{A}}  = \hat{B} + [\hat{A},\hat{B}] + \frac{1}{2!} [\hat{A},[\hat{A},\hat{B}]] + \dots,
\end{equation}
and the commutation relations of $su(2)$: $[\hat{S}^z, \hat{S}^+] = \hat{S}^+,\ [\hat{S}^z, \hat{S}^-] = -\hat{S}^-,\ [\hat{S}^+, \hat{S}^-] =2 \hat{S}^z$.
Finally,
\begin{align}
\left( \frac{\partial \hat{U}_j^s}{\partial \xi_j^-} \right) (\hat{U}_j^s)^{-1} 
= e^{-\xi_j^z} \left( \hat S_j^- + 2\xi_j^+ \hat S_j^z - \xi_j^{+2} \hat S_j^+ \right).
\end{align}
Equating the coefficients of each $\hat S_j^a$ one obtains
\begin{subequations}\label{eq:disentanglementConditions}
\begin{align}
i \Phi_j^+ &= \dot{\xi}_j^+ - e^{-\xi_j^z} \xi_j^{+2} \dot{\xi}_j^-  - \xi_j^+ \dot{\xi_j^z} ,  \\ 
i \Phi_j^z &= \dot{\xi_j^z} + 2 \xi_j^+ e^{-\xi_j^z} \dot{\xi_j^-}, \\
i \Phi_j^- &= e^{-\xi_j^z} \dot{\xi_j^-}.
\end{align}
\end{subequations}
Rearranging for $\dot\xi_j^a$ yields the SDEs \cite{ringelGritsev}
\begin{subequations}
\begin{align}
-i \dot{\xi}^+_j &= \Phi^+_j + \Phi^z_j \xi^+_j - \Phi^-_j {\xi^+_j}^2 ,  \\
-i \dot{\xi}^z_j &= \Phi^z_j - 2 \Phi^-_j \xi^+_j , \\
-i \dot{\xi}^-_j &= \Phi^-_j \exp{\xi^z_j}.
\end{align}
\end{subequations}

\section{Ito and Stratonovich Conventions}\label{app:itoStratonovich}
In order to consistently define a stochastic differential equation, it
is necessary to specify a discretization convention
\cite{kloeden}. 
These are distinguished by how the values of a function $\bar{f}(t_j)$ defined at discrete times $t_j\equiv j \Delta t$ are assigned from the values of its continuous counterpart $f(t)$. Different discretization
schemes are parameterized by a constant $\alpha$ as
\begin{equation}
\bar{f}(t_j)=\alpha f(t_j)+(1-\alpha) f(t_{j-1}), \quad 0 \leq \alpha \leq 1.
\end{equation}
The choice $\alpha =0$ gives the Ito convention $\bar{f}(t_j) =f(t_{j-1})$, while $\alpha=1/2$ gives the Stratonovich convention. The latter corresponds to choosing $\bar{f}(t_j)$ as the average of the values of $f(t)$ at $t_{j-1}$, $t_j$. Since $[f(t_{j}) + f(t_{j-1})]/2 \approx f([t_{j} + t_{j-1}]/2)$, the Stratonovich convention is also known as the \textit{mid-point} prescription. SDEs in the Stratonovich convention satisfy the rules of ordinary
calculus. However, when working with Ito SDEs a specific calculus is
required \cite{ito,kloeden}.  If we interpret the disentangling
equations (\ref{eq:disentanglementGeneral}) as SDEs, they are to be understood as initially expressed in the Stratonovich convention. This is the form which arises naturally in physical applications involving well-defined continuous processes, i.e. noise with a finite
  correlation time, in the limit of the correlation time going to zero. However, equations in the Ito convention are typically
mathematically and computationally simpler to handle. It is therefore
often convenient to translate Stratonovich SDEs into the Ito form.  In
the Stratonovich form, the SDE for the disentangling
variables $\xi_i^a$, collectively represented as a vector $\xi_S$,
can be written as
\begin{equation}
\frac{\mathrm{d}\xi_S}{\mathrm{d}t} =A_S(\xi_S,t)+B_S(\xi_S,t)\phi,
\end{equation}
where $\phi$ is a vector composed of the stochastic variables
$\phi^a_j$, $A_S$ is the drift vector and $B_S$ is a matrix of diffusion coefficients. The corresponding SDE for the vector $\xi$ in the Ito
convention is given by
\begin{equation}\label{eq:ito}
\frac{\mathrm{d}\xi}{\mathrm{d}t} =A(\xi,t)+B(\xi,t)\phi ,
\end{equation}
where
\begin{subequations}
\begin{align}\label{eq:itoStrat}
A &=A_S + \frac{1}{2}  (B^T \nabla_{\xi}) B^T ,  \\
B&=B_S.
\end{align} 
\end{subequations}
For the quantum Ising model~(\ref{eq:isingH}), this modification only
affects the Ito SDE~(\ref{eq:xpSDE}) for $\xi^+_j$, which becomes
\begin{equation}\label{eq:xiIto}
-i \dot{\xi}^+_j = \frac{1}{2}\Gamma (1- {\xi^+_j}^2)+ \frac{1}{2} \xi^+_j \sum_k O_{jk} O_{jk}+ \xi^+_j \sum_k O_{jk} \phi_k /\sqrt{i} .
\end{equation}
In many cases, the extra term $\xi^+_j \sum_k O_{jk} O_{jk}/2$
vanishes, since $O O^T = 2 \mathcal{J}$ and the interaction matrix
$\mathcal{J}$ typically has no diagonal elements.  However, as
discussed in Appendix \ref{app:diag}, for system sizes $N$ that are
multiples of $4$ we add a diagonal constant shift to $\mathcal{J}$, in
order to make it invertible. In this case, the SDE for $\xi^+_j$ takes
the form~(\ref{eq:xiIto}). This leads to different behavior for the
classical variables, but does not affect the resulting physical
observables.

\section{Analytical Averaging of the Equations of Motion}\label{app:analyticalAveraging}
As we discussed in Section \ref{subsec:Ito}, in principle it is
possible to analytically average the SDEs governing the dynamics of
physical observables. The expectation value of an observable
$\hat{\mathcal{O}}$ following time-evolution from an initial state
$\ket{\psi_0}$ can be expressed as
\begin{equation}\label{eq:obsApp}
\langle \hat{\mathcal{O}}(t) \rangle = \langle f (t)\rangle_{\phi, \tilde{\phi}}, 
\end{equation}
where
\begin{align}\label{eq:definition_f}
f(t) = \bra{\psi_0}[\hat{U}^s(\tilde{\xi}(t))]^\dag \hat{\mathcal{O}} \hat{U}^s(\xi(t)) \ket{\psi_0}. 
\end{align}
Here, $\hat{U}^s= \otimes_i \hat{U}^s_i$ and the two time-evolution
operators depend on independent stochastic processes $\phi$ and
$\tilde{\phi}$ via $\xi[\phi]$ and $\tilde{\xi}[\tilde{\phi}]$. The
functional form of $f$, in terms of the disentangling variables $\xi$
and $\tilde{\xi}$, depends on the chosen observable and the initial
state. The equation of motion of $f$ is obtained from the Ito chain
rule as given by Eq.~(\ref{eq:EoMs}) in the main text. This can be
written as $\dot {f}= \Upsilon f$, where we define the linear operator
\begin{equation}
  \label{eq:EoMs_appendix}
  \Upsilon\equiv \sum_{ia} (A^a_i + \sum_{jb} B^{ab}_{ij} \phi^b_j) \frac{\partial }{\partial \xi^a_i} + \frac{1}{2} \sum_{ijab} \sum_{ck} B^{ac}_{ik}  B^{bc}_{jk}  \frac{\partial^2}{\partial\xi_i^a \partial\xi_j^b}.
 \end{equation}
For notational economy, the indices $a$ and $b$ run over $\{ +, -, z
\}$ and over both the $\xi,\tilde{\xi}$ variables. The analytical
expression for the average $\langle d \mathcal{O}(t)/dt \rangle
=\langle \dot{f} \rangle_{\phi,\tilde{\phi}}$ can be obtained by applying~(\ref{eq:EoMs_appendix}) to the definition~(\ref{eq:definition_f})  and averaging over the HS fields $\phi$, $\tilde{\phi}$:
\begin{align}\label{eq:differentiate_f}
 \begin{split}
\langle \dot{f} \rangle_{\phi, \tilde{\phi}} = \, & \Big\langle \bra{\psi_0} \left(\Upsilon[\hat{U}^s(\tilde{\xi})]^\dag \right) \hat{\mathcal{O}} \hat{U}^s(\xi) \ket{\psi_0} \Big\rangle_{\phi, \tilde{\phi}}\\
& \, +   \Big\langle \bra{\psi_0}[\hat{U}^s(\tilde{\xi})]^\dag \hat{\mathcal{O}} \left( \Upsilon \hat{U}^s(\xi) \right) \ket{\psi_0} \Big\rangle_{\phi, \tilde{\phi}} .
\end{split}
\end{align}
Since $\langle\hat{U}^s\rangle_{\phi} = \hat{U}$, we can simplify
(\ref{eq:differentiate_f}) using
\begin{align}\label{eq:tEvolApp}
\left\langle \Upsilon \hat{U}^s(\xi)  \right\rangle_{\phi} =\left\langle  \frac{d}{dt} \hat{U}^s(\xi) \right\rangle_{\phi }   = -i \hat{H} \hat{U}.
\end{align} 
This can also be verified by directly evaluating $\Upsilon \hat{U}^s(t)$ and using the commutation relations of $su(2)$. Similarly,
\begin{align}\label{eq:tEvolApp}
\left\langle \Upsilon [\hat{U}^s(\tilde{\xi})]^\dag  \right\rangle_{ \tilde{\phi}} = \left\langle  \frac{d}{dt} [\hat{U}^s(\tilde{\xi})]^\dag \right\rangle_{\tilde{\phi} }   = i  \hat{U}^\dag \hat{H} .
\end{align} 
Using these identities, the equation of motion
(\ref{eq:differentiate_f}) for $\langle \hat{\mathcal{O}}(t) \rangle$
can be written as
\begin{align}\label{eq:heisenbergEoM}
\bra{ \psi_0 } \frac{d\hat{\mathcal{O}}}{dt}  \ket{\psi_0} = \langle \dot{f} \rangle_{\phi, \tilde{\phi}}  =  i  \bra{ \psi_0 }  \left( \hat{H}  \hat{U}^\dag \hat{\mathcal{O}}  \hat{U}  -  \hat{U}^{ \dagger } \hat{\mathcal{O}} \hat{U}  \hat{H}  \right) \ket{\psi_0}. 
\end{align}
This can be recognized as a matrix element of the Heisenberg equation
of motion
\begin{equation}\label{eq:heisenbergEoMgen}
\frac{d}{dt}\hat{\mathcal{O}}(t) = i [\hat{H},\hat{\mathcal{O}}(t)] .
\end{equation} 
Such matrix elements give rise to a set of coupled first order ODEs,
whose number in general grows exponentially with the system size.
Solving these ODEs is therefore equivalent to diagonalizing the
Hamiltonian.
\section{Building Blocks for Local Observables}
\label{app:buildingBlocks} As discussed in Section \ref{sec:stochastic}, expectation values of
products of local operators, starting from a product state, can be
expressed in terms of stochastic averages over products of on-site ``building
blocks''. This follows from the fact that the time-evolution operator
can be factorized over lattice sites as $U(t)=\langle
U^s(t)\rangle_\phi$ where $\hat{U}^s(t) \equiv \otimes_i
\hat{U}_i^s(t)$. For example, if there are no spin operators in the
observable $\hat{\mathcal O}$ at site $i$ we may use the building
blocks
\begin{widetext}
\begin{subequations}\label{eq:normalizations}
\begin{align}
\begin{split}
{\mathcal B}^{\uparrow\uparrow}_i&\equiv {}_i\bra{\uparrow} \hat{U}^{s}_i(\tilde{\xi})^\dag \hat{U}^s_i(\xi) \ket{\uparrow}_i\\  & =  e^{-\frac{\xi^z_i +\tilde{\xi}^{z*}_i}{2}} \left( e^{\xi^z_i +\tilde{\xi}^{z*}_i} + e^{\tilde{\xi}^{z*}_i}\xi^-_i \xi^+_i + e^{\xi^z_i} \tilde{\xi}^{-*}_i \tilde{\xi}^+_i + \xi^-_i \tilde{\xi}^{-*}_i (1+\xi^+_i \tilde{\xi}^{+*}_i) \right) , \end{split} \\
{\mathcal B}^{\uparrow\downarrow}_i&\equiv{}_i \bra{\uparrow}  \hat{U}^{s}_i (\tilde{\xi})^\dag \hat{U}^s_i(\xi) \ket{\downarrow}_i = e^{-\frac{\xi^z_i +\tilde{\xi}^{z*}_i}{2}} \left( \tilde{\xi}^{-*}_i + e^{\tilde{\xi}_i^{z*}} \xi^+_i + \tilde{\xi}^{-*}_i \xi^+_i \tilde{\xi}^{+*}_i  \right) , \\
{\mathcal B}^{\downarrow\uparrow}_i&\equiv {}_i \bra{\downarrow}  \hat{U}^{s}_i(\tilde{\xi})^\dag  \hat{U}^s_i(\xi) \ket{\uparrow}_i = e^{-\frac{\xi^z_i +\tilde{\xi}^{z*}_i}{2}} \left( \xi^-_i + e^{\xi_i^z} \tilde{\xi}^{+*}_i + \xi^-_i \xi^+_i \tilde{\xi}^{+*}_i  \right) ,  \\
{\mathcal B}^{\downarrow\downarrow}_i&\equiv {}_i \bra{\downarrow}  \hat{U}^{s}_i(\tilde{\xi})^\dag  \hat{U}^s_i(\xi) \ket{\downarrow}_i = e^{-\frac{\xi^z_i +\tilde{\xi}^{z*}_i}{2}} ( 1 +  \xi^+_i \tilde{\xi}^{+*}_i ),
\end{align}
\end{subequations}
\end{widetext}
depending on the initial and final states. If the spin operator
$S_i^z$ is present in $\hat{\mathcal O}$ we may use one of the
following:
\begin{widetext}
\begin{subequations}\label{eq:szs}
\begin{align}
{\mathcal B}^{z\uparrow\uparrow}_i& \equiv{}_i \bra{\uparrow}  \hat{U}^{s}_i(\tilde{\xi})^\dag \hat{S}^z_i \hat{U}^s_i(\xi) \ket{\uparrow}_i   =  \frac{1}{2} e^{-\frac{\xi^z_i +\tilde{\xi}^{z*}_i}{2}} \left( e^{\xi^z_i +\tilde{\xi}^{z*}_i} + e^{\tilde{\xi}^{z*}_i}\xi^-_i \xi^+_i + e^{\xi^z_i} \tilde{\xi}^{-*}_i \tilde{\xi}^{+*}_i + \xi^-_i \tilde{\xi}^{-*}_i (-1+\xi^+_i \tilde{\xi}^{+*}_i) \right) ,   \\
{\mathcal B}^{z\uparrow\downarrow}_i&\equiv{}_i \bra{\uparrow}  \hat{U}^{s}_i(\tilde{\xi})^\dag \hat{S}^z_i \hat{U}^s_i(\xi) \ket{\downarrow}_i = \frac{1}{2} e^{-\frac{\xi^z_i +\tilde{\xi}^{z*}_i}{2}} \left( - \tilde{\xi}^{-*}_i + e^{\tilde{\xi}_i^{z*}} \xi^+_i + \tilde{\xi}^{-*}_i \xi^+_i \tilde{\xi}^{+*}_i  \right) ,  \\
{\mathcal B}^{z\downarrow\uparrow}_i&\equiv {}_i \bra{\downarrow}  \hat{U}^{s}_i(\tilde{\xi})^\dag \hat{S}^z_i \hat{U}^s_i(\xi) \ket{\uparrow}_i =\frac{1}{2} e^{-\frac{\xi^z_i +\tilde{\xi}^{z*}_i}{2}} \left( -\xi^-_i + e^{\xi_i^z} \tilde{\xi}^{+*}_i + \xi^-_i \xi^+_i \tilde{\xi}^{+*}_i  \right) , \\
{\mathcal B}^{z\downarrow\downarrow}_i&\equiv {}_i \bra{\downarrow}  \hat{U}^{s}_i(\tilde{\xi})^\dag \hat{S}^z_i  \hat{U}^s_i(\xi) \ket{\downarrow}_i = \frac{1}{2} e^{-\frac{\xi^z_i +\tilde{\xi}^{z*}_i}{2}} ( - 1 +  \xi^+_i \tilde{\xi}^{+*}_i ) .
\end{align}
\end{subequations}
\end{widetext}
Similarly, if the observable $\hat{\mathcal O}$ contains $\hat{S}^+_i$, $\hat{S}^-_i$ we may use:
\begin{widetext}
\begin{subequations}\label{eq:sps}
\begin{align}
{\mathcal B}^{+\uparrow\uparrow}_i& \equiv{}_i \bra{\uparrow}  \hat{U}^{s}_i(\tilde{\xi})^\dag \hat{S}^+_i \hat{U}^s_i(\xi) \ket{\uparrow}_i = e^{-\frac{{\tilde{\xi}^{z}_i}+{\xi^{z*}_i}}{2}} {\xi^-_i}  \left(({\tilde{\xi}^-_i} {\tilde{\xi}^+_i})^*+e^{{\tilde{\xi}^{z*}_i}}\right) 
 \\
{\mathcal B}^{+\uparrow\downarrow}_i&\equiv{}_i \bra{\uparrow}  \hat{U}^{s}_i(\tilde{\xi})^\dag \hat{S}^+_i \hat{U}^s_i(\xi) \ket{\downarrow}_i = e^{-\frac{{\tilde{\xi}^{z}_i}+{\xi^{z*}_i}}{2}} \left(({\tilde{\xi}^-_i} {\tilde{\xi}^+_i})^*+e^{{\tilde{\xi}^{z*}_i}}\right) ,  \\
{\mathcal B}^{+\downarrow\uparrow}_i&\equiv {}_i \bra{\downarrow}  \hat{U}^{s}_i(\tilde{\xi})^\dag \hat{S}^+_i \hat{U}^s_i(\xi) \ket{\uparrow}_i = e^{-\frac{{\tilde{\xi}^{z}_i}+{\xi^{z*}_i}}{2}} {\xi^-_i} {\tilde{\xi}^{+*}_i}  , \\
{\mathcal B}^{+\downarrow\downarrow}_i&\equiv {}_i \bra{\downarrow}  \hat{U}^{s}_i(\tilde{\xi})^\dag \hat{S}^+_i  \hat{U}^s_i(\xi) \ket{\downarrow}_i =e^{-\frac{{\tilde{\xi}^{z}_i}+{\xi^{z*}_i}}{2}} {\tilde{\xi}^{+*}_i}.
\end{align}
\end{subequations}
\begin{subequations}\label{eq:sms}
\begin{align}
{\mathcal B}^{-\uparrow\uparrow}_i& \equiv{}_i \bra{\uparrow}  \hat{U}^{s}_i(\tilde{\xi})^\dag \hat{S}^-_i \hat{U}^s_i(\xi) \ket{\uparrow}_i  =e^{-\frac{{\tilde{\xi}^{z}_i}+{\xi^{z*}_i}}{2}} {\tilde{\xi}^{-*}_i}  \left({\xi^-_i} {\xi^+_i}+e^{{\xi^z_i}}\right)  ,   \\
{\mathcal B}^{-\uparrow\downarrow}_i&\equiv{}_i \bra{\uparrow}  \hat{U}^{s}_i(\tilde{\xi})^\dag \hat{S}^-_i \hat{U}^s_i(\xi) \ket{\downarrow}_i = e^{-\frac{{\tilde{\xi}^{z}_i}+{\xi^{z*}_i}}{2}} {\xi^+_i} {\tilde{\xi}^{-*}_i} ,  \\
{\mathcal B}^{-\downarrow\uparrow}_i&\equiv {}_i \bra{\downarrow}  \hat{U}^{s}_i(\tilde{\xi})^\dag \hat{S}^-_i \hat{U}^s_i(\xi) \ket{\uparrow}_i =e^{-\frac{{\tilde{\xi}^{z}_i}+{\xi^{z*}_i}}{2}} \left({\xi^-_i} {\xi^+_i}+e^{{\xi^z_i}}\right) , \\
{\mathcal B}^{-\downarrow\downarrow}_i&\equiv {}_i \bra{\downarrow}  \hat{U}^{s}_i(\tilde{\xi})^\dag \hat{S}^-_i  \hat{U}^s_i(\xi) \ket{\downarrow}_i =e^{-\frac{{\tilde{\xi}^{z}_i}+{\xi^{z*}_i}}{2}} {\xi^+_i} .
\end{align}
\end{subequations}
\end{widetext}
For example, starting in the fully-polarized initial state
$\ket{\Downarrow}$ one readily obtains
\begin{equation}
  \langle \hat S_i^x(t)\rangle =\frac{1}{2}\langle \hat S_i^++\hat S_i^-\rangle=\frac{1}{2}
  \langle ({\mathcal B}_i^{+\downarrow\downarrow}+{\mathcal B}_i^{-\downarrow\downarrow})\prod_{j\neq i}{\mathcal B}_j^{\downarrow\downarrow}\rangle_{\phi,\tilde\phi}.
\end{equation}
Using the above results one arrives at the exact formula
\begin{equation}
\langle \hat{S}^x_i(t) \rangle =\frac{1}{2} \Big\langle e^{-\sum_j  \left( \frac{\xi^z_j+\tilde{\xi}^{z*}_j}{2} \right) } (\xi^+_i + \tilde{\xi}^{+*}_i)  \prod_{j\neq i} (1+\xi^+_j \tilde{\xi}^{+*}_j)\Big\rangle_{\phi,\tilde{\phi}}. 
  \end{equation}
Similarly,
\begin{equation}
\langle \hat{S}^y_i(t) \rangle =  \frac{i}{2} \Big\langle e^{-\sum_j  \left( \frac{\xi^z_j+\tilde{\xi}^{z*}_j}{2} \right) } (\xi^+_i - \tilde{\xi}^{+*}_i)  \prod_{j\neq i} (1+\xi^+_j \tilde{\xi}^{+*}_j)\Big\rangle_{\phi,\tilde{\phi}}.
\end{equation}
The result for $\langle \hat S_i^z(t)\rangle$ is given by
(\ref{eq:generalMagnetization}) in the main text.
\section{Ising Stochastic Differential Equations}\label{app:isingSDEs}
As we discussed in Section \ref{sec:ising}, the Ito SDEs for the
quantum Ising model are given by Eq.~(\ref{eq:isingSDEs}),
which we repeat here for convenience:
\begin{subequations}\label{eq:RTIsing}
\begin{align}
-i \dot{\xi}^+_j &= \frac{\Gamma}{2} (1-{\xi^+_j}^2) + \xi^+_j \sum_k
O_{jk} \phi_k/\sqrt{i} \label{eq:xpIsing}, \\ -i \dot{\xi}^z_j &=
-\Gamma \xi^+_j + \sum_k O_{jk} \phi_k /\sqrt{i} \label{eq:xzIsing},
\\ - i \dot{\xi}^-_j &= \frac{\Gamma}{2}\exp{\xi^z_j}.\label{eq:xmIsing}
\end{align}
\end{subequations}
It is readily seen that the disentangling variable $\xi_j^+$ plays a
particularly important role for the quantum dynamics in this
parameterization: as we will discuss, $\xi_j^+$ vanishes identically
in the classical limit $\Gamma=0$, and it is the only disentangling
variable that is not dependent on the others, as follows from
(\ref{eq:xpIsing}). Once $\xi_j^+$ is known, $\xi_j^z$ can be obtained
by integrating (\ref{eq:xzIsing}) with respect to time. In turn,
$\xi_j^-$ has a deterministic dependence on $\xi_j^z$, as follows from
(\ref{eq:xmIsing}).
The non-linearity of the equation of motion (\ref{eq:xpIsing}) for
$\xi_j^+$, renders it non-trivial to solve. However, exact solutions
to the full set of SDEs (\ref{eq:RTIsing}) are readily obtained in the
classical limit with $\Gamma=0$, and in the non-interacting limit with
$J=0$. We consider each below. 

In the classical limit with $\Gamma=0$, the equation of motion for
$\xi^+_j(t)$ becomes linear. Due to the initial condition
$\xi^+_j(0)=0$ one obtains the trivial solution
$\xi^+_j(t)=0$. Similarly, $\xi^-_j(t)=0$. The variable $\xi^z_j(t)$
undergoes Brownian motion and its time-evolution can be computed as
\begin{equation}\label{eq:xzClassical}
\xi_j^z(t) = i \sum_k O_{jk} \int^t_0 \mathrm{d}t^\prime \phi_k(t^\prime) =  \sqrt{i} \sum_k O_{jk} W_k(t)  .
\end{equation}
The quantities $W_k(t)$ are a set of $N$ independent standard Wiener processes, which can be numerically generated as
\begin{equation}\label{eq:generateWiener}
W_k(t) = \sqrt{t} \mathcal{N}(0,1) ,
\end{equation}
where $\mathcal{N}(0,1)$ is a random number extracted from a
zero-mean, unit-variance Gaussian distribution. In the classical limit
with $\Gamma=0$, there is no dynamics; this result can be recovered by
substituting (\ref{eq:xzClassical}) into the stochastic expressions
for observables. For example, for the Loschmidt amplitude
(\ref{eq:loschGen}) and the magnetization (obtained using the building
blocks in Appendix \ref{app:buildingBlocks}) one  obtains
\begin{align}
|A(t)| &= 1, \label{eq:A1}\\ \langle \hat{S}^z_j(t) \rangle&= \langle
\hat{S}^z_j(0) \rangle, \label{eq:m1}
\end{align}
for any initial condition. For example, for quantum quenches from the
fully-polarized initial state $\ket{\Downarrow}$, the Loschmidt amplitude and magnetization can be obtained by substituting~(\ref{eq:xzClassical}) and $\xi^{+}_i=0$ into (\ref{eq:survivalAmplitudeStochastic}) and (\ref{eq:generalMagnetization}):
\begin{align}\label{eq:classicalEvolution}
A(t) & = \Big\langle  \exp\left( \frac{(i+1)}{2} \sqrt{NJ} W_1(t) \right) \Big\rangle_\phi, \\
\begin{split}
\mathcal{M}(t) & =  - \frac{1}{2}\Big\langle  \exp\left(\frac{\sqrt{NJ}}{2}[ (1+i)W_1(t) +  (1-i) \tilde{W}_1(t)] \right) \Big\rangle_{\phi \tilde{\phi}} ,
\end{split}
\end{align}
where $W_1(t)$ and $\tilde{W}_1(t)$ are independent Wiener processes
obtained as in Eq.~(\ref{eq:generateWiener}). Averaging the above
equations with respect to the noises $\phi$, one obtains the results
(\ref{eq:A1}) and (\ref{eq:m1}). From (\ref{eq:xzClassical}), one
can also calculate the variance of the stochastic functions $f_A$ and
$f_\mathcal{M}$ corresponding to these observables, via $A(t)\equiv
\langle f_A \rangle_{\phi}$ and $\mathcal{M}(t)\equiv \langle
f_\mathcal{M} \rangle_{\phi, \bar{\phi}}$. One obtains
   \begin{align}\label{eq:varLosch}
\sigma^2(f_A) &= e^{\frac{NtJ}{2}} -1 , \\
\sigma^2(f_\mathcal{M}) &=\frac{1}{4}\left( e^{NtJ} -1 \right) .
\end{align}
In both cases, the variance grows exponentially with time and the
system size. The similar functional form of the Loschmidt
  amplitude~(\ref{eq:loschGen}) and the
  magnetization~(\ref{eq:generalMagnetization}), which both involve exponential factors of $e^{-\sum_i \xi^z_i(t)/2}$, leads to similar behavior for the fluctuations. An extra factor of two is present in
  the exponent for the magnetization due to the presence of two
  Hubbard--Stratonovich transformations for local observables. The
  presence of the exponential factors is suggestive of the exponential
  growth of fluctuations even for non-zero $\Gamma$. This is observed
  numerically and is discussed in the main text.

In the non-interacting limit with $J=0$ the equations of motion for the disentangling variables become deterministic, as $O_{jk}=0$. These
can be solved analytically:
\begin{subequations}\label{eq:rtNint}
\begin{align}
\xi^+_j(t)&= i \tan(\Gamma t/2) \label{eq:xpNI} , \\
{\xi}^z_j(t) &=  - 2 \log \cos (\Gamma t/2)  \label{eq:xzNI},  \\
{\xi}^-_j(t) &= i \tan ( \Gamma t /2).
\end{align}
\end{subequations}
As expected, (\ref{eq:rtNint}) parameterizes the precession of a
single spin in a magnetic field applied along the $x$-direction. This
can be seen by writing the time-evolved state
$\ket{\psi(t)}=\hat{U}(t)\vert_{J=0}\ket{\psi(0)}$ for product-state
initial conditions $\ket{\psi(0)} = \otimes_j \left(a_j
\ket{\uparrow}_j + b_j \ket{\downarrow}_j \right)$ with $|a_j|^2 +
|b_j|^2=1$, using the values of $\xi$ given in (\ref{eq:rtNint}). This
yields
\begin{equation}
\ket{\psi(t)}=   \bigotimes_j \begin{pmatrix}
a_j \cos(\Gamma t/2)  - i b_j \sin(\Gamma t/2) \\ -i a_j \sin(\Gamma t/2) + b_j \cos(\Gamma t/2)
  \end{pmatrix}.
\end{equation}

\section{Moments of the Disentangling Variables}
\label{app:moments}
As we noted in Section \ref{sec:correlationsClassical}, certain
  averages of the classical disentangling variables are identically
  zero for all times. Here, we go further and demonstrate that a
    set of monomials in $R_i \equiv \mathrm{Re} (\xi^+_i)$ and $I_i
  \equiv \mathrm{Im}(\xi^+_i)$ have vanishing averages for all $t$.
To see this we note that the coupled SDEs for $R_i$ and $I_i$ can be
obtained by  combining Eq.~(\ref{eq:xpIsing})
with its complex conjugate:
\begin{subequations}\label{eq:realImagEvol}
\begin{align}
\dot{R}_i(t)=& \Gamma R_i I_i +\frac{\sqrt{2}}{2} [R_i( O_{Ri}  - O_{Ii}) - I_i( O_{Ri}+ O_{Ii})] \phi \\
\begin{split}
\dot{I}_i(t)=& \frac{\Gamma}{2}(1- R_i^2 + I_i^2) \\ &+ \frac{\sqrt{2}}{2}[R_i (O_{Ri}  + O_{Ii}) + I_i(O_{Ri}- O_{Ii}) ]\phi 
\end{split}
\end{align}
\end{subequations}
In the above, we have introduced the shorthand notation $O_{Ri}\phi \equiv \sum_j (O_R)_{ij} \phi_j$, where $(O_R)_{ij} \equiv \mathrm{Re}(O_{ij})$,
and similarly for $O_I$.  
The identities we wish to show are most easily proved by introducing a
convenient formal notation; this makes it simpler to analyze the
system of ODEs which arise from averaging the SDEs. In particular, let
us represent the classical average of a given monomial $\langle
R_i^n I_i^m \rangle_\phi $ as a \textit{state}
$\ket{n_i,m_i}$. To compute the time-evolution of this state, one
  applies the Ito chain rule~(\ref{eq:EoMs}); this requires
  differentiating with respect to $R_i$ and $I_i$. Due to (\ref{eq:realImagEvol}), each differentiation with respect to
$R_i$ ($I_i$) decreases $n_i$ ($m_i$) by $1$, and annihilates a state
where $n_i$ ($m_i$) is equal to zero. This suggests that we can
formally represent derivatives as \textit{annihilation
  operators} $\hat a_{Ri} \equiv \frac{\partial}{\partial R_i} $, $\hat a_{Ii}
\equiv \frac{\partial}{\partial I_i}$ satisfying
\begin{subequations}
\begin{align}
\hat a_{Ri} \ket{n,m} = n_i \ket{n_i-1 ,m_i}  , \\
\hat a_{Ii} \ket{n,m} = m_i \ket{n_i ,m_i-1} .
\end{align}
\end{subequations}
Following the same line of reasoning, we can represent $R_i$ and $I_i$ themselves as \textit{creation operators} satisfying
\begin{subequations}
\begin{align}
\hat a^\dag_{Ri} \ket{n_i,m_i} = \ket{n_i+1 ,m_i}  , \\
\hat a^\dag_{Ii} \ket{n_i,m_i} = \ket{n_i ,m_i+1} .
\end{align}
\end{subequations}
It can be seen that the operators satisfy bosonic commutation
relations $[\hat a_X, \hat a^\dag_{X^\prime}]=\delta_{XX^\prime}$
where $X,X^\prime \in \{R_i,I_i\}$. We can use the notations
introduced above to compactly write the equation of motion for a
general state $\ket{n_i,m_i} = \langle R_i^n I_i^m \rangle_\phi$, obtained
via the Ito chain rule, as
\begin{equation}\label{eq:RTschr}
\frac{\mathrm{d}}{\mathrm{d}t} \ket{n_i,m_i}  = - \hat H_i\ket{n_i,m_i} 
\end{equation}
where we have defined an effective Hamiltonian $\hat H_i$:
\begin{equation}
\begin{split}\label{eq:RThamilt}
& \hat H_i \equiv \ \Gamma \hat a_{Ri}^\dag \hat a_{Ii}^\dag \hat a_{Ri} + \frac{\Gamma}{2} \left(1-\hat a^\dag_{Ri} \hat a^\dag_{Ri} +\hat a_{Ii}^\dag \hat a^\dag_{Ii} \right) \hat a_{Ii}  \\
& \, + \frac{1}{2} (O_I O_I^T  + O_R O_R^T)_{ii} (\hat a^\dag_{Ri} \hat a^\dag_{Ri}+ \hat a^\dag_{Ii} \hat a^\dag_{Ii}) (\hat a_{Ri} \hat a_{Ri}+\hat a_{Ii} \hat a_{Ii} ).
\end{split}
\end{equation}
In deriving Eq.~(\ref{eq:RThamilt}), we used the Ito calculus property
$\langle f(t) \phi(t)\rangle_\phi=0$, and the
properties of $O_R$ and $O_I$ given in Appendix~\ref{app:diag}. In
Eq.~(\ref{eq:RTschr}), the time-evolution of a general classical average $\langle R^n_i(t) I_i^m(t)\rangle_\phi$ has been cast in a form that is reminiscent of a Euclidean Schr\"odinger equation,
where the real-time variable $t$ plays the role of an imaginary-time
variable.
The notations introduced above allow us to obtain an infinite number
of identities for monomials in $R_i$, $I_i$. As it can be seen from~(\ref{eq:RThamilt}), the Hamiltonian does not contain any term
that raises or lowers the index $n_i$ of a state $\ket{n_i,m_i}$ by an
odd number. Hence, states with odd and even $n_i$ belong to separate
even and odd subspaces $\mathcal{H}_{e}$ and $\mathcal{H}_{o}$. At
$t=0$, the initial condition $\xi^+_i(0)=0$ translates into the
initial conditions $\ket{n_i,m_i}=\delta_{n_i0}\delta_{m_i0}$. Since
$\ket{0,0} \equiv \langle 1 \rangle_\phi$ does not belong to the odd
subspace $\mathcal{H}_{o}$, all states $\ket{n_i,m_i} \in
\mathcal{H}_{o}$ vanish identically at all times. This finding can be
expressed as
\begin{align}\label{eq:vanishing}
\langle R_i^n(t) I_i^m(t) \rangle_\phi &=0, \quad \forall \quad n \ \mathrm{ odd} .
\end{align}
This provides an analytical proof of the numerical observation that
Re$\langle \xi^+_i(t)\rangle_\phi$ vanishes at all times, as we
previously reported in Ref.~\cite{stochasticApproach}.  Thus far, we
have focused on moments involving variables at a single site $i$. The
multi-site generalization is straightforward and is analogous to the
construction of a many-site Fock space from a collection of
single-site ones. Creation and annihilation operators are defined
analogously to the same-site case; from their definition, it is clear
that operators acting at different sites commute, since they
  are just multiplications or differentiations by independent
  variables. Of particular interest is the case of monomials
involving two sites $i \neq j$, associated with two times $t_i$,
$t_j$. States can be defined as
\begin{align}\label{eq:2sitestate}
\begin{split}
  \langle R_i^{n_i}(t_i) I_i^{m_i}(t_i)  R_j^{n_j}(t_j) I_j^{m_j}(t_j) \rangle_\phi  \equiv \\
  \Big|(n_i,m_i,t_i)_i;(n_j,m_j,t_j)_j\Big\rangle.
\end{split}
\end{align}
Consider, for example, the evolution of the state with respect to $t_i$. For $t_i \neq t_j$, this is given by Eq.~(\ref{eq:RTschr})
where the state is replaced by~(\ref{eq:2sitestate}). For $t_j=t_i$,
the effective Hamiltonian becomes
\begin{align}\label{eq:RThamilt2}
\begin{split}
& \hat H_2 = \hat H_i + \hat H_j \\
&\, +\frac{1}{2}(O_I O_I^T  + O_R O_R^T )_{ij} (\hat a^\dag_{Ri} \hat a^\dag_{Rj}+ \hat a^\dag_{Ii} \hat a^\dag_{Ij} ) (\hat a_{Ri} \hat a_{Rj}+\hat a_{Ii} \hat a_{Ij})\\
&\, +\frac{1}{2} (O_I O_I^T  + O_R O_R^T )_{ij} (\hat a^\dag_{Ri} \hat a^\dag_{Ij} -\hat a^\dag_{Ii} \hat a^\dag_{Rj}) (\hat a_{Ri}\hat a_{Ij}-\hat a_{Ii} \hat a_{Rj} ) \\
&\, +\mathcal{J}_{ij} (\hat a^\dag_{Ri} \hat a^\dag_{Rj} -\hat a^\dag_{Ii} \hat a^\dag_{Ij}) (\hat a_{Ri} \hat a_{Ij}+\hat a_{Ii} \hat a_{Rj}) \\
&\, -\mathcal{J}_{ij} (\hat a^\dag_{Ri} \hat a^\dag_{Ij}+ \hat a^\dag_{Ii} \hat a^\dag_{Rj}) (\hat a_{Ri} \hat a_{Rj}-\hat a_{Ii} \hat a_{Ij}) .
\end{split} 
\end{align} Eq.~(\ref{eq:RThamilt2}) allows us to generalize the result~(\ref{eq:vanishing}) for monomials involving two sites. Consider the total number of $R$ factors in a given monomial, $n=n_i+n_j$. When $i$ and $j$ are not nearest neighbors, the last two terms in~(\ref{eq:RThamilt2}) vanish, and both the Hamiltonians~(\ref{eq:RThamilt}) and~(\ref{eq:RThamilt2}) conserve the overall evenness or oddness of $n$. By the same argument as in the one-site case, states with odd $n$ must therefore vanish for all $t_i$, $t_j$.
This finding can be used to prove the vanishing of ${\rm Im}\,C^{zz}_n(t)$ with $n \geq 2$, as defined in Eq.~(\ref{eq:correlationsClassical}) and numerically studied in Section~\ref{sec:classical}. Using Ito calculus, we find that $C^{zz}_n(t)$ is given by
\begin{widetext}
\begin{align}\label{eq:CzzConn}
C^{zz}_n(t) =  -\Gamma \sum_{i} \int\displaylimits_0^t \hspace{-2mm} \int\displaylimits_0^t \mathrm{d}t_1  \mathrm{d}t_2 \left[  \langle \xi^+_i(t_1)  \xi^+_{i+n}(t_2) \rangle_\phi -   \langle \xi^+_i(t_1) \rangle_\phi \langle  \xi^+_{i+n}(t_2) \rangle_\phi \right].
\end{align}
\end{widetext}
By Eq.~(\ref{eq:vanishing}), we know that $\langle \xi^+_i(t)\rangle$ is purely imaginary for all $i$, $t$, such that the second contribution to Eq.~(\ref{eq:CzzConn}) is real-valued. The first contribution involves
\begin{align}
\begin{split}
\langle \xi^+_i(t_1) \xi^+_{i+n}(t_2) \rangle_\phi & =  \langle R_i(t_1) R_{i+n}(t_2) \rangle_\phi - \langle I_i(t_1) I_{i+n}(t_2) \rangle_\phi \\
& +i \langle R_i(t_1) I_{i+n}(t_2) \rangle_\phi + i \langle I_i(t_1) R_{i+n}(t_2) \rangle_\phi .
\end{split} 
\end{align}
It can be seen that the imaginary part of $\langle \xi^+_i(t_1)
\xi^+_{i+n}(t_2)\rangle_\phi$ features only monomials with odd $n$,
which vanish for any $t_1$, $t_2$ by the previous argument.  Thus, the
connected correlation function $C^{zz}_n(t)$ with $n \geq 2$ must be
real valued.  The operator description of the evolution of moments
introduced in this Section allows us to identify vanishing expectation
values in a transparent way, without solving the
SDEs~(\ref{eq:RTIsing}). The presence of identically vanishing moments
suggests that the Ising SDEs may contain a degree of redundancy, and
that it may be possible to reduce them to a simpler form which
automatically takes into account these vanishing averages.  The
operator formalism employed to find the vanishing moments provides a
rather general alternative viewpoint, which may turn out to be a
useful tool for future developments.

\end{document}